\documentclass[epj]{svjour}
\usepackage{graphicx}

\usepackage{rotating}
\usepackage[T1]{fontenc}
\usepackage[latin1]{inputenc}
\usepackage{psfrag}
\def\lapprox{{\raise0.5ex\hbox{$<$}\hskip-0.80em\lower0.5ex\hbox{$\sim$}

}}

\def\gapprox{{\raise0.5ex\hbox{$>$}\hskip-0.80em\lower0.5ex\hbox{$\sim$}

}}

\usepackage{graphics}
\begin{document}

\title{Two-Pion Production in Proton-Proton Collisions with Polarized Beam}  

\author{
S.~Abd El-Bary\inst{8}\and
S.~Abd El-Samad\inst{8}\and
R.~Bilger\inst{6}\and
K.-Th.~Brinkmann\inst{10}\and
H.~Clement\inst{6} \and
M.~Dietrich\inst{6}\and
E.~Doroshkevich\inst{6}\and
S. Dshemuchadse\inst{5,2}\and
A.~Erhardt\inst{6}\and
W.~Eyrich\inst{3}\and
A.~Filippi\inst{7}\and
H.~Freiesleben\inst{2}\and
M.~Fritsch\inst{3,1}\and
R.~Geyer\inst{4}\and
A.~Gillitzer\inst{4}\and
C.~Hanhart\inst{4}\and
J.~Hauffe\inst{3}\and
K.~Haug\inst{6}\and
D.~Hesselbarth\inst{4}\and
R.~Jaekel\inst{2}\and
B.~Jakob\inst{2}\and
L.~Karsch\inst{2}\and
K.~Kilian\inst{4}\and
H.~Koch\inst{1}\and
J. Kress\inst{6}\and
E.~Kuhlmann\inst{2}\and
S.~Marcello\inst{7}\and
S.~Marwinski\inst{4}\and
R.~Meier\inst{6}\and
K. M\"oller\inst{5}\and
H.P. Morsch\inst{4}\and
L.~Naumann\inst{5}\and
J.~Ritman\inst{4}\and
E.~Roderburg\inst{4}\and
P. Sch\"onmeier\inst{2,3}\and
M. Schulte-Wissermann\inst{2}\and
W.~Schroeder\inst{3}\and
M.~Steinke\inst{1}\and
F. Stinzing\inst{3}\and
G.Y. Sun\inst{2}\and
J.~W\"achter\inst{3}\and
G.J.~Wagner\inst{6}\and
M.~Wagner\inst{3}\and
U.~Weidlich\inst{6}\and
A. Wilms\inst{1}\and
P.~Wintz\inst{8}\and
S.~Wirth\inst{3}\and
G.~Zhang\inst{6}\thanks{present address: Peking University}\and
P. Zupranski\inst{9}
}
%
\mail{H. Clement \\email: clement@pit.physik.uni-tuebingen.de}
%

\institute{
Ruhr-Universit\"at Bochum, Germany \and
Technische Universit\"at Dresden, Germany \and
Friedrich-Alexander-Universit\"at Erlangen-N\"urnberg, Germany \and
Forschungszentrum J\"ulich, Germany \and
Forschungszentrum Rossendorf, Germany \and
Physikalisches Institut der Universit\"at T\"ubingen, Auf der Morgenstelle 14,
D-72076 T\"ubingen, Germany \and
University of Torino and INFN, Sezione di Torino, Italy \and
Atomic Energy Authority NRC Cairo, Egypt \and
Soltan Institute for Nuclear Studies, Warsaw, Poland\and
Rheinische Friedrich-Wilhelms Universit\"at Bonn, Germany 
\\
(COSY-TOF Collaboration)}
\date{\today}
%
%
\abstract{ The two-pion production reaction $\vec{p}p\rightarrow
  pp\pi^+\pi^-$ was measured with a
 polarized proton beam at $T_p \approx$ 750 and 800 MeV using the
  short version of the COSY-TOF spectrometer. The implementation of a
  delayed pulse technique for Quirl and central calorimeter provided
  positive $\pi^+$ identification in addition to the standard particle
  identification, energy determination as well as time-of-flight and
  angle measurements.  Thus all four-momenta of the emerging particles
  could be determined with 1-4 overconstraints.  Total and differential cross
  sections as well 
  as angular distributions of the vector analyzing power have been
  obtained. They are compared to previous data and theoretical
  calculations. In contrast to predictions we find significant
  analyzing power values up to $A_y$ = 0.3.  }

\maketitle
\section{Introduction}
\label{intro}

The $\Delta(1232)$ resonance is the most dominant resonance in the
pion-nucleon system. Therefore pion production in nucleon-nucleon
collisions is strongly affected by its excitation if kinematically and
dynamically allowed~\cite{EW,delta_juel}. This is not the case in
two-pion production near threshold. Since the $\Delta$ resonance
decays by emission of only one pion, single-$\Delta$ excitation can
contribute to two-pion production only if a second pion is produced
associatedly. However, near the two-pion production threshold it is
very unlikely that the second nucleon involved in the collision
process gets excited, too. Hence the associated production of the
second pion can only proceed by non-resonant rescattering, which however, is a
very weak process as calculated by Ref. \cite{luis}.

This situation provides a unique opportunity to study excitation and decay
of the Roper resonance $N^*(1440)$, because this process provides the only
possibility for resonant two-pion production at energies close to the two-pion
threshold - as has been predicted in detailed theoretical calculations
\cite{luis} and as has been demonstrated by the first exclusive measurements
performed by the PROMICE/ \\
WASA collaboration at the CELSIUS ring in the energy
range of $T_p$ = 650 - 775 MeV \cite{WB,JP,JJ}.

The measured angular distributions in the center-of-mass system (cms) 
are all consistent with isotropy with the
exception of the proton angular distribution, which is shaped by the
dominating $\sigma$ exchange between the two colliding protons - as
demonstrated in Ref. \cite{WB} and predicted  by Ref. \cite{luis}. From this we
learn that except for the  

\onecolumn{
\begin{figure}[H]
\begin{center}
\includegraphics[width=44pc]{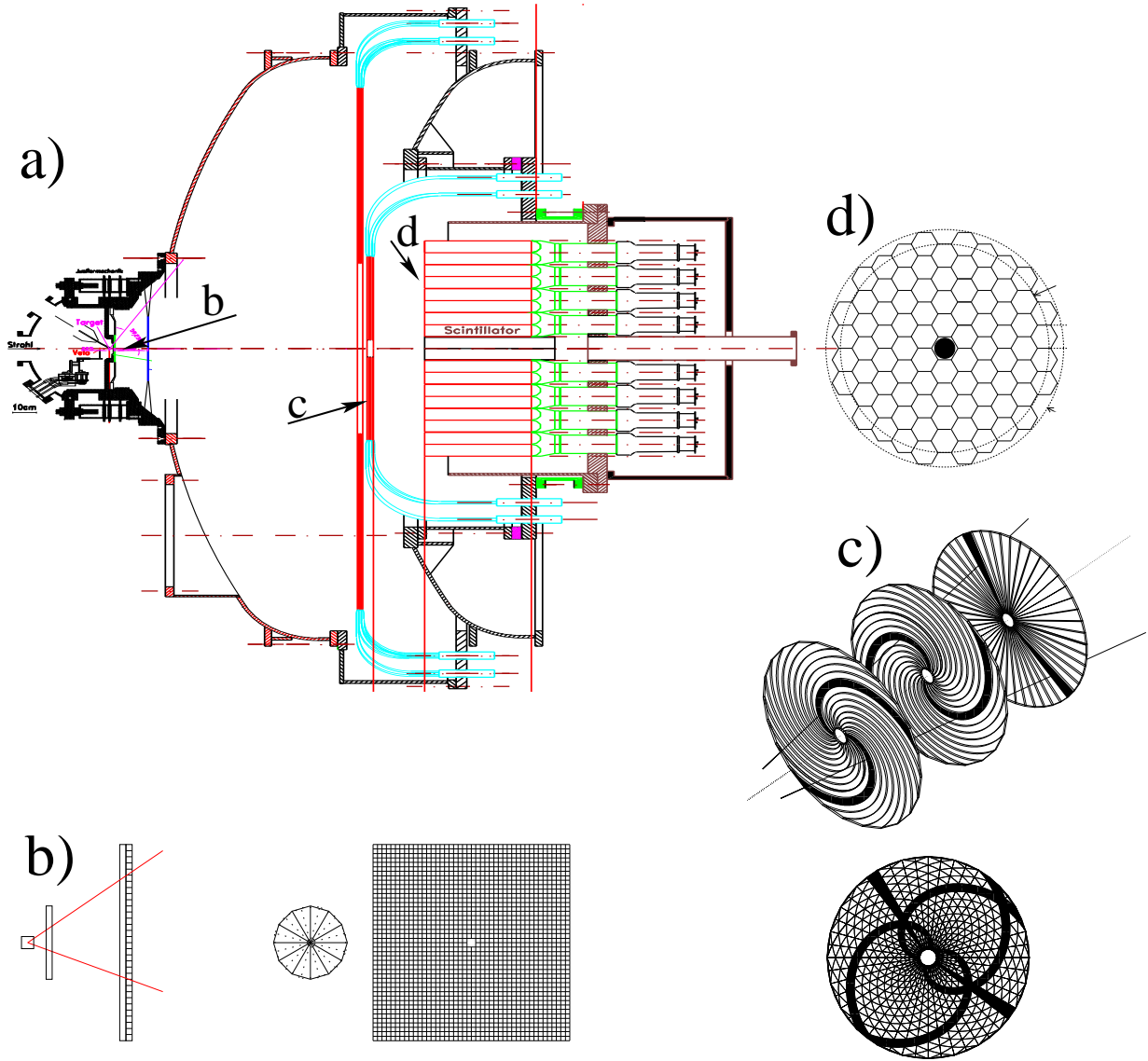}
\end{center}
\caption{Sketch of the short COSY-TOF setup used in this experiment showing
  (a) the full detector arrangement with inserts for (b) the start detector
  region with start wedges and hodoscope displaying both side (left) and front
  (right) views, (c) the central hodoscope ("Quirl")
  as the stop detector for TOF measurements and (d) the central
  calorimeter. The positions of the subdetector systems (b) - (d) in the full
  setup are indicated in (a). For a detailed description see text. } 
\end{figure}
}

\twocolumn{ relative motion of the two protons all ejectiles are
in relative s-waves to each other at beam energies close to threshold. This is
not surprising, since the leading mechanism for 
two-pion production, which is the excitation of the Roper resonance and its
subsequent decay $N^* \to N\sigma$ into
  the $\sigma$ channel $(\pi\pi)_{I=l=0}$, produces only particles in
  relative s-waves. Also the Roper decay via the $\Delta$
  resonance $N^* \to \Delta\pi \to N\pi^+\pi^-$ produces predominantly
  only relative s-waves, although both pions are produced in a p-wave.
  Since no evidence of any significant p-wave contribution (except partially
  between the two emitted protons) is observed in the differential
  cross section, the dominating partial wave
  for two-pion production near threshold must be the $^1S_0$ partial
  wave in the entrance channel.  However, for non-zero analyzing
  powers at least two interfering partial waves are necessary.  Hence,
  from the Roper excitation and decay we do not expect sizable
  non-zero analyzing power values.  Indeed, the calculations of Ref.
  \cite{luis}, which include also the two-pion production via the
  $\Delta\Delta$ system, predict vanishing analyzing powers. This
  result, however, should be taken with some caution, since the
  calculations have been carried out without taking into account
  initial state interactions,
  which should be a prime source of imaginary parts of the amplitudes
  and analyzing powers are proportional to the imaginary part of interference
  terms.

On the other hand analyzing powers originate from the interference of
amplitudes. Hence small amplitudes in the production process, in particular
p-wave amplitudes, can be easily sensed in polarized beam measurements

\section{Experiment}
\label{sec:2}

The measurements have been carried out at the J\"ulich Cooler Synchrotron COSY
 at beam energies of nominally 747 and 793 MeV using the time-of-flight
 spectrometer TOF at an external beam
line. The  setup of the TOF detector system is displayed in Fig. 1. At
the entrance of the detector system the beam - focussed to a diameter smaller
than 2 mm - hits the LH$_2$ target, which has a length of 4 mm, a diameter of
6 mm and 0.9 $\mu m$ thick hostaphan foils as entrance and exit windows
\cite{has}. At a distance of 22 mm downstream of the target the two layers of
the start detector (each consisting of 1 mm thick scintillators cut into 12
wedge-shaped sectors) were placed followed by a two-plane fiber hodoscope
(96 fibers per plane, 2 mm thick fibers ) at a distance of 165 mm from target,
see Fig. 1b. Whereas the 
start detector mainly supplies the start times for the time-of-flight (TOF)
measurements, the fiber hodoscope primarily provides a good angular resolution
for the detected particle tracks. In its central part the TOF-stop detector
system consists of the so-called Quirl, a 3-layer scintillator system  1081 mm 
downstream of the target shown in Fig. 1c and described in detail in
Ref. \cite{dah} -  and in its peripheral part of the so-called Ring, also a
3-layer scintillator system built in a design analogous to the Quirl, however,
with inner 
and outer radii of 560 and 1540 mm, respectively. Finally behind the Quirl a
calorimeter (Fig. 1a,d) was installed for identification of charged
particles and of neutrons as well as for measuring the energy of charged
particles. The calorimeter, details of which are given in Ref. \cite{kress},
consists of 84 hexagon-shaped scintillator blocks of length 450 mm, which
suffices to stop deuterons, protons and pions of energies up to 400, 300 and
160 MeV, respectively.

The particle identification is done with the standard $\Delta$E-E technique
using the energy signal from the calorimeter together with the TOF-signal of
the Quirl as shown in Fig. 2 and as described in more detail in
Ref.\cite{evd}, where also sample 
spectra are shown. 

In order to distinguish $\pi^+$ and $\pi^-$ particles we
have implemented the delayed-pulse technique into the calorimeter setup. By
using  multihit TDCs the decay of a $\mu^+$ resulting from
the decay of a $\pi^+$ stopped in one of the calorimeter blocks can be
registered by an afterpulse arriving in the $\mu$s range after the prompt
pulse. In order to prevent false afterpulses to be registered, which stem from
prompt pulses arising from follow-up events, we use the last layer of the
Quirl as veto for afterpulses. This technique for $\pi^+$ identification had
previously been installed at the forward detector of PROMICE/WASA and
CELSIUS/WASA, respectively, where it has been shown to work reliably
\cite{hans,janusz}. Fig. 3 shows a sample spectrum of registered delayed pulses
at a proton beam of $T_p$ = 793 MeV. The spectrum is very clean following the
expected exponential decay pattern of $\mu^+$ in matter with a lifetime of 2.2
$\mu$s. The overall efficiency of this $\pi^+$ identification has been
determined to be 39 $\%$. The efficiency losses mainly stem from  up-stream
moving positrons, which originate from $\mu^+$ decay in the calorimeter and
create a veto signal in the last layer of the Quirl.

\begin{figure}
\begin{center}
\includegraphics[width=10pc]{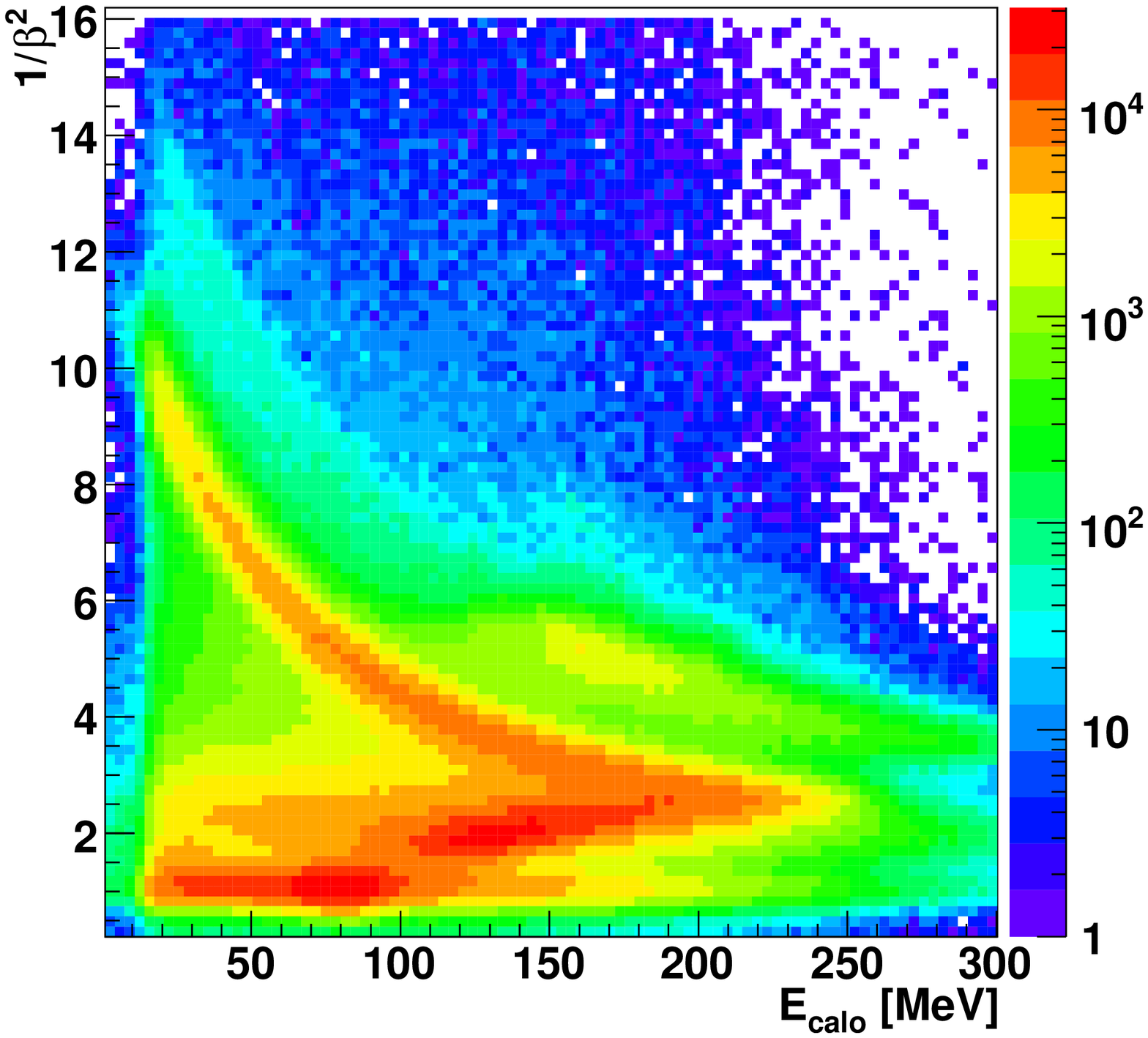}
\includegraphics[width=10pc]{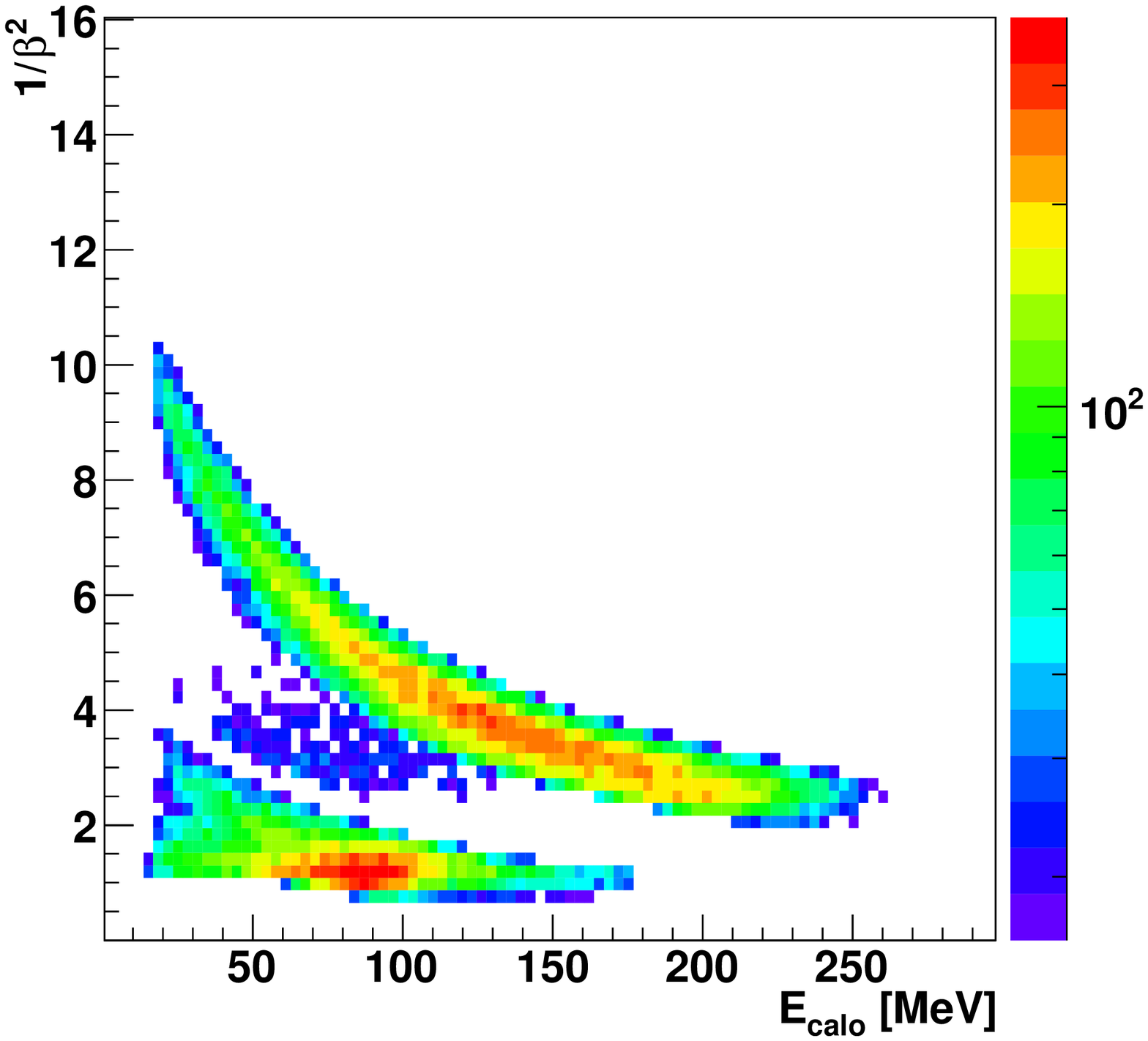}
\end{center}
\caption{$\Delta E - E$ plots for particles hitting the calorimeter. The
  $\Delta E$ information is taken from the TOF information obtained by Quirl
  and start detector and plotted as $1 / \beta^2$, where $\beta = v / c$
  denotes the particle velocity relative to c. For the $E$ information the
  energy deposited in the calorimeter has been taken. Left: without any
  trigger selections and conditions. Right: with trigger selection for three-
  and four-prong events as well as with missing mass conditions $MM_{pp} \ge$
  260 MeV and 100 MeV $\le MM_{pp\pi} \le$ 180 MeV, see text.   
  }
\end{figure}

In the experiment the trigger for two-pion production was set to at least
three hits in the Quirl-Ring system associated with two hits in the start
detector.That way single-pion production events could be excluded, which at
most can give two hits in the Quirl-Ring system. For calibration and
monitoring purposes, however, a two-prong trigger was also set up with a high
prescale factor.
From straight-line fits to the hit detector elements the tracks of charged
particles were reconstructed. They are accepted as good tracks, if they
originate in the target and have a hit in each detector element the track
passes through. In this way the angular resolution is better than 1$^\circ$
both in azimuthal and in polar angles. 
By construction of the calorimeter a particle may hit
one or more calorimeter blocks. The number of blocks hit by a particular
particle is given by the track reconstruction. The total energy deposited by
this particle in the calorimeter is then just the (calibrated) sum
of energies deposited in all blocks belonging to the particular track.

In order to have maximum angular coverage by the detector elements and to
minimize the fraction of charged pions decaying in flight before reaching the
stop detectors, the short version of the TOF spectrometer
was used. In this way a total polar angle coverage of 3$^\circ \leq
\Theta^{lab}\leq$ 49$^\circ$ was achieved with the central calorimeter
covering the region 3$^\circ \leq\Theta^{lab}\leq$ 28$^\circ$. 

By identifying and reconstructing three or four charged tracks of an event the
reaction channel $pp\pi^+\pi^-$ can be fully identified and reconstructed
kinematically complete with 1 - 4 overconstraints. 
The maximum  possible laboratory (lab) polar angle for protons
emerging from two-pion production events is $\approx 30^\circ$. Hence
practically all protons are within the 
angular acceptance of the calorimeter and can be identified  as protons
there. For charged pions emerging from two-pion production events there is no
kinematic limit at the incident beam energies of interest. Hence the angular
coverage has not been complete for their detection with this setup. However,
since detection of two protons and one pion are enough to safely reconstruct
the second pion of an event and since due to the identity of the initial
collision partners the unpolarized angular distributions are symmetric about
90$^\circ$ in the overall center-of-mass system (cms), still most of the
reaction phase space has been covered by this experiment. The total acceptance
has 
been 15 $\%$ of the full phase space, which is a factor 5 -7 larger than
in the PROMICE/WASA measurements, where the angular acceptance was $\leq$
21$^\circ$.
 
Though most of the events stemming from single-pion production could be
removed by the three- and four-prong trigger conditions, there is still
appreciable background left from $pp\pi^0$ events, where the $\pi^0$ either
undergoes Dalitz decay or where the $\gamma$s emerging from $\pi^0$ decay
undergo conversion in  matter. In both cases $e^+e^-$ pairs are produced,
which easily may be misidentified as charged pions. And since the $pp\pi^0$
production cross 
section is three orders of magnitude larger than that for $pp\pi^+\pi^-$ this
background is sizeable. In order to suppress this background we applied two
constraints on missing mass spectra as shown in Fig.4. For the missing mass
$MM_{pp\pi}$ of two identified protons and one identified pion, which has to
be equivalent to the mass of the second pion in case of a true two-pion
production event, we require 100 MeV $\leq MM_{pp\pi} \leq$ 180 MeV. We see
from Fig. 4 that by this condition the $MM_{pp}$ spectrum gets already very
clean. To be on the safe side we introduce as further constraint $MM_{pp}
\geq$ 260 MeV before applying the kinematic fit to events, who have passed
both these conditions. For the kinematic fits we have one overconstraint
in case of three-prong $pp\pi$ events and 3 overconstraints in case of
four-prong events, where all emerging particles have been recorded in the
detector.

\begin{figure}
\begin{center}
\includegraphics[width=15pc]{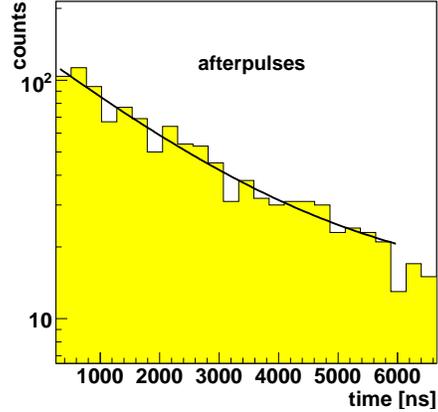}
\end{center}
\caption{Time spectrum of afterpulses detected by the delayed pulse technique
  installed at Quirl and Calorimeter.The logarithmic slope given by the solid
  curve corresponds to $\tau$ = 2.18(6) $\mu$s, which agrees very well with the
  muon lifetime. In the fit a small background of 10 counts per bin  (1
  bin corresponds to 256 ns) is assumed.  } 
\end{figure}

\begin{figure}
\begin{center}
\includegraphics[width=10pc]{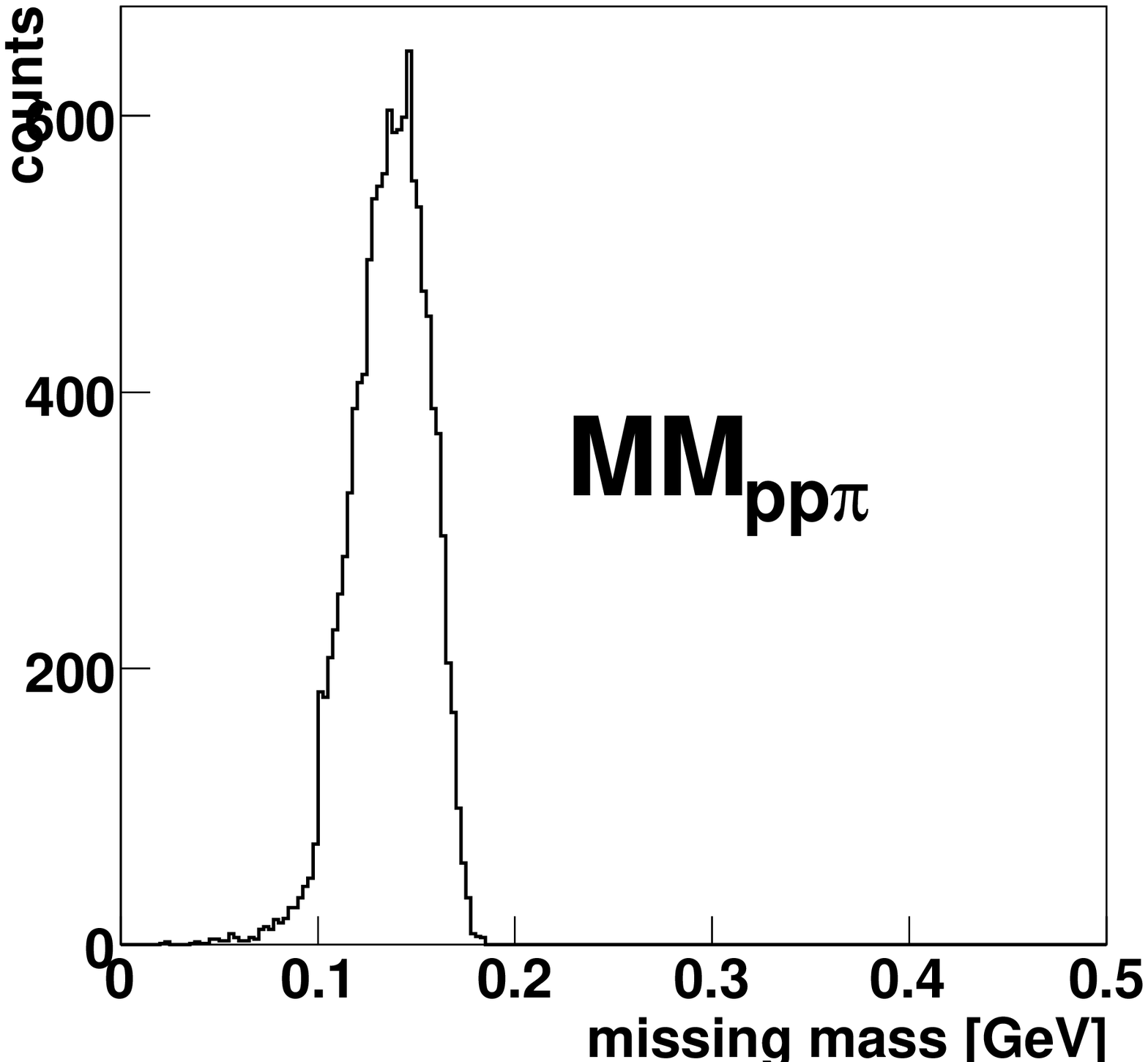}
\includegraphics[width=10pc]{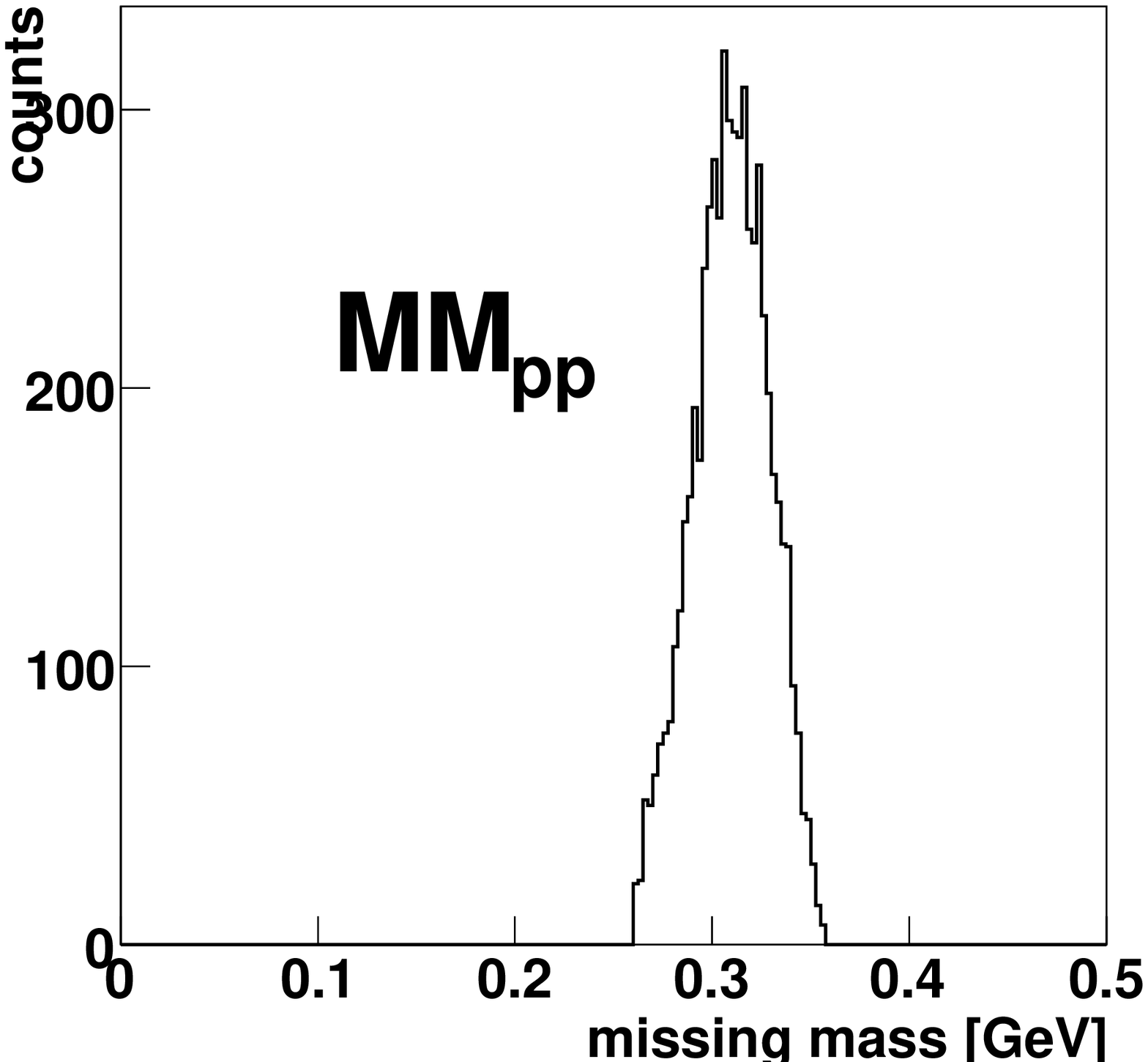}
\end{center}
\caption{Left: missing mass spectrum $MM_{pp\pi}$ of two identified protons and
  one identified pion with the condition $MM_{pp} \geq$ 260 MeV. Right:
  missing mass spectrum $MM_{pp}$ of two identified protons with the condition
  100 MeV $\leq MM_{pp\pi} \leq$ 180 MeV.} 
\end{figure}

The luminosity of the experiment was determined as  $dL/dt =
5*10^{28}~cm^{-2}s^{-1}$ from the analysis of $pp$
elastic scattering, which was measured simultaneously utilizing a prescaled
two-prong trigger for hits in the Ring detector. Due to their opening angle of
$\delta_{pp}\approx80^\circ$ 
between both tracks, such two-track events have both hits in the Ring. They
are easily identified by using in addition the coplanarity constraint
$170^\circ < \Delta\Phi \leq 180^\circ$, i.e. using the same procedure as
described in more detail in Ref. \cite{evd}. Adjustment of the $pp$-elastic
data in absolute height to the SAID values 
gives the required luminosity.
All data have been efficiency corrected by MC
simulations of the detector setup by using the CERN GEANT3 \cite{geant}
detector simulation 
package, which accounts both for electromagnetic and hadronic interactions of
the ejectiles with the detector materials. 


The measurements have been conducted with a vector-polarized proton beam
providing a steady change of "spin-up" and "spin-down" runs in subsequent
spills of about 10~s duration. "Spin-up" is defined of having the spin of the
protons perpendicular to the plane defined by the COSY accelerator ring and 
pointing to the sky
(y-axis). According to the Madison convention \cite{madison} we then use a
coordinate system, where the z-axis is along the beam direction and the
x-axis points to the horizontal left corresponding to the azimuthal angle of
$\Phi$ = 0$^\circ$ being at 9 o'clock when looking downstream the beam
direction. 

For the determination of the beam polarisation at the target we use again 
elastic scattering, determine its countrate asymmetries and compare them to
the known analyzing powers of elastic scattering as compiled in the SAID
database \cite{SAID}. The countrate asymmetry $\epsilon (\Theta,\Phi)$ at a specific solid
angle $(\Theta$, $\Phi)$
for a certain particle of an event is given by the difference of countrates at
"spin-up"$N_{\uparrow}(\Theta, \Phi)$ and "spin-down" $N_{\downarrow}(\Theta,
\Phi)$ normalized to their sum:
\begin{center}
$\epsilon (\Theta,\Phi)$ = $\frac{N_{\uparrow}(\Theta, \Phi) -
N_{\downarrow}(\Theta,\Phi)}{N_{\uparrow}(\Theta, \Phi) +
N_{\downarrow}(\Theta,\Phi)} $ = $ P_y * A_y(\Theta,\Phi),~~~~~~~(1) $ \
\end{center}
where $P_y$ is the beam polarisation and 
\begin{center}
$A_y(\Theta,\Phi) = A_y(\Theta) * cos(\Phi)~~~~~~~~~~~~~~~~~~~~~~~~~~~~~~(2) $ 
\end{center}
is the analyzing power of the reaction of interest with the particle of
interest detected at the solid angle $(\Theta$, $\Phi)$.

In order to determine and also to monitor online the beam polarisation we have
selected  
pp elastic scattering events with 32$^\circ \leq \Theta_{lab} \leq $
33$^\circ$, where the SAID database \cite{SAID} gives $A_y$ = 0.30. A sample of 
the measured countrate asymmetry $\epsilon (\Theta,\Phi)$ is given in Fig. 5
exhibiting the expected cosine dependence on the azimuthal angle. A fit to the
data with eqs.(1) and (2) yields the beam
polarisation $P_y$. During the beamtime the beam polarisation was
very stable within a few percents and with very high values of typically $P_y$
= 0.87.


\begin{figure}
\begin{center}
\includegraphics[width=15pc]{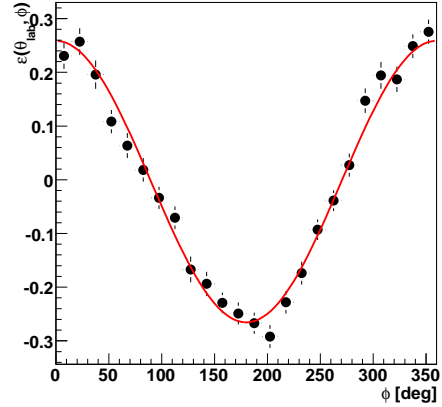}
\end{center}
\caption{Countrate asymmetry $\epsilon (\Theta_{lab}, \Phi)$ in dependence of
  the azimuthal angle $\Phi$ for elastic scattering events measured in the
  polar angle interval 32$^\circ \leq \Theta_{lab} \leq $
33$^\circ$. The solid curve is a fit to the data with a $cos(\Phi)$
dependence.} 
\end{figure}

\begin{figure}
\begin{center}
\includegraphics[width=15pc]{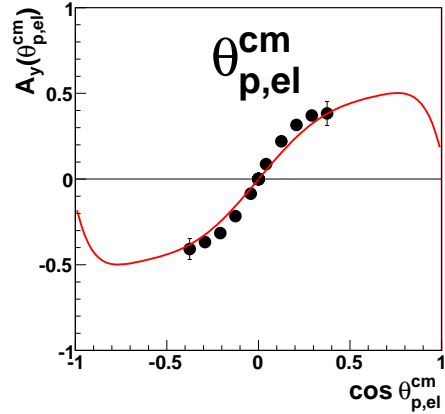}
\end{center}
\caption{Analyzing power of elastic scattering events measured in the Ring
  detector in comparison with the SAID database \cite{SAID} (solid line). }
\end{figure}

Due to its cylinder symmetry the TOF detector is ideally
suited for polarisation measurements. Thus, the analyzing power $A_y$
may be determined experimentally by use of the super ratio $R(\Theta,
\Phi)$ of the four countrates for the combinations of "spin-up",
spin-down" with "left" or "right" :
\begin{center}
$R(\Theta,\Phi) = \sqrt{\frac{N_{\uparrow}(\Theta, \Phi) *
    N_{\downarrow}(\Theta, \Phi + 
  180^\circ)}{ N_{\downarrow}(\Theta, \Phi) *  N_{\uparrow}(\Theta, \Phi +
  180^\circ)}}~~~~~~~~~~~~~~~~~~~(3)$   
\end{center}
and 
\begin{center}
$A_y(\Theta,\Phi) = \frac{1}{P_y} * \frac{R(\Theta,\Phi) - 1}{R(\Theta,\Phi) +
  1} ~~~~~~~~~~~~~~~~~~~~~~~~~~~~~~~(4)$ 
\end{center}

In this superratio systematic uncertainties, detector efficiencies etc. cancel
to a very large extent making thus the deduced analyzing power essentially
free of systematic asymmetries, which often are not under control in "single
arm" measurements with a not cylinder-symmetric detector.
By using the four countrates we in addition may define an "anti-analyzing
power" (instrumental asymmetry) $\bar{A_y}$ with the alternative super ratio $\bar{R}$ defined as 

\begin{center}
$\bar{R}(\Theta,\Phi) = \sqrt{\frac{N_{\uparrow}(\Theta, \Phi) *
    N_{\uparrow}(\Theta, \Phi + 180^\circ)}{ N_{\downarrow}(\Theta, \Phi) *
    N_{\downarrow}(\Theta, \Phi + 180^\circ)}},~~~~~~~~~~~~~~~~~~~(5)$
\end{center}
which leads to $\bar{A_y}$ = 0 for all cases with the exception of systematic
changes in beam direction when flipping the spin of the projectiles. Hence
$\bar{A_y}$ serves as a control observable to uncover systematic asymmetries
of such kind.

Due to their trivial cos$\Phi$ dependence the analyzing powers are usually
quoted as
\begin{center}
$A_y(\Theta) = A_y(\Theta, \Phi$ = 0). 
\end{center}
In
order to take into account the fully available statistics, we have sorted the
data into $\Theta$ and $\Phi$ bins, calculated  $A_y(\Theta,\Phi)$ for each of
these bins and fitted for each particular $\Theta$ bin the $\Phi$ dependence
of $A_y(\Theta,\Phi)$ by a cosine dependence. As a result we obtain for each
$\Theta$ bin an $A_y(\Theta)$ value, which contains the full statistics of the
measurements over the full  $\Phi$ range.

\begin{figure}
\begin{center}
\includegraphics[width=22pc]{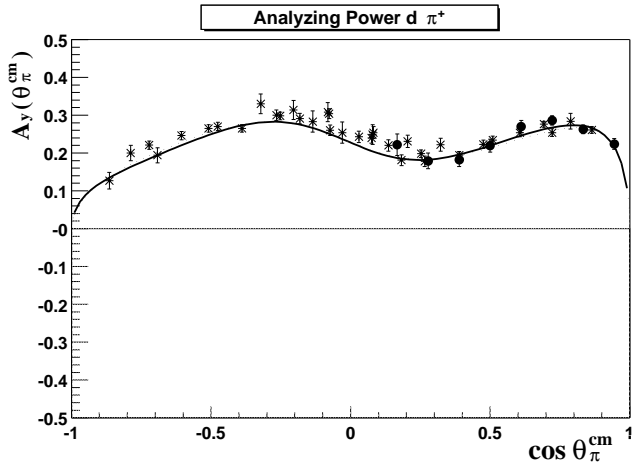}
\end{center}
\caption{Analyzing power data for the $pp \to d \pi^+$ reaction as measured in
  the TOF detector (solid circles) in comparison with the SAID database
  \cite{SAID}(solid line) and 
  LAMPF data \cite{tipp} (asterisk symbols). }
\end{figure}

As an example we show in Figs. 6 and 8 analyzing power and anti-analyzing power
(instrumnetal asymmetry) distributions obtained for $pp$ elastic scattering events recorded in the Ring
detector. The analyzing power data agree well with SAID in the
measured range. As a further check of the 
deduced beam polarisation we have evaluated also the $pp \to d \pi^+$ reaction
utilising the same 2-prong trigger as used for the selection of elastic
scattering events. The results for the selected $d \pi^+$ events is shown in
Fig. 7 in comparison with the SAID data base \cite{SAID} and LAMPF data
\cite{tipp}. Again there is good agreement.

The anti-analyzing power (intsrumental asymmetry) data are all consistent with zero
giving thus no indication of systematic asymmetries in the measurements. As a
typical example of an anti-analyzing power distribution obtained from
two-pion production events we show in Fig.8, right, the distribution according
to the 
angle $\Theta_{\pi}^{cm}$ of any of the pions in the overall cms. Again the
data are consistent with zero, whereas  $A_y(\Theta_\pi^{cm}) \neq$ 0, see
Fig. 15.

\begin{figure}
\begin{center}
\includegraphics[width=10pc]{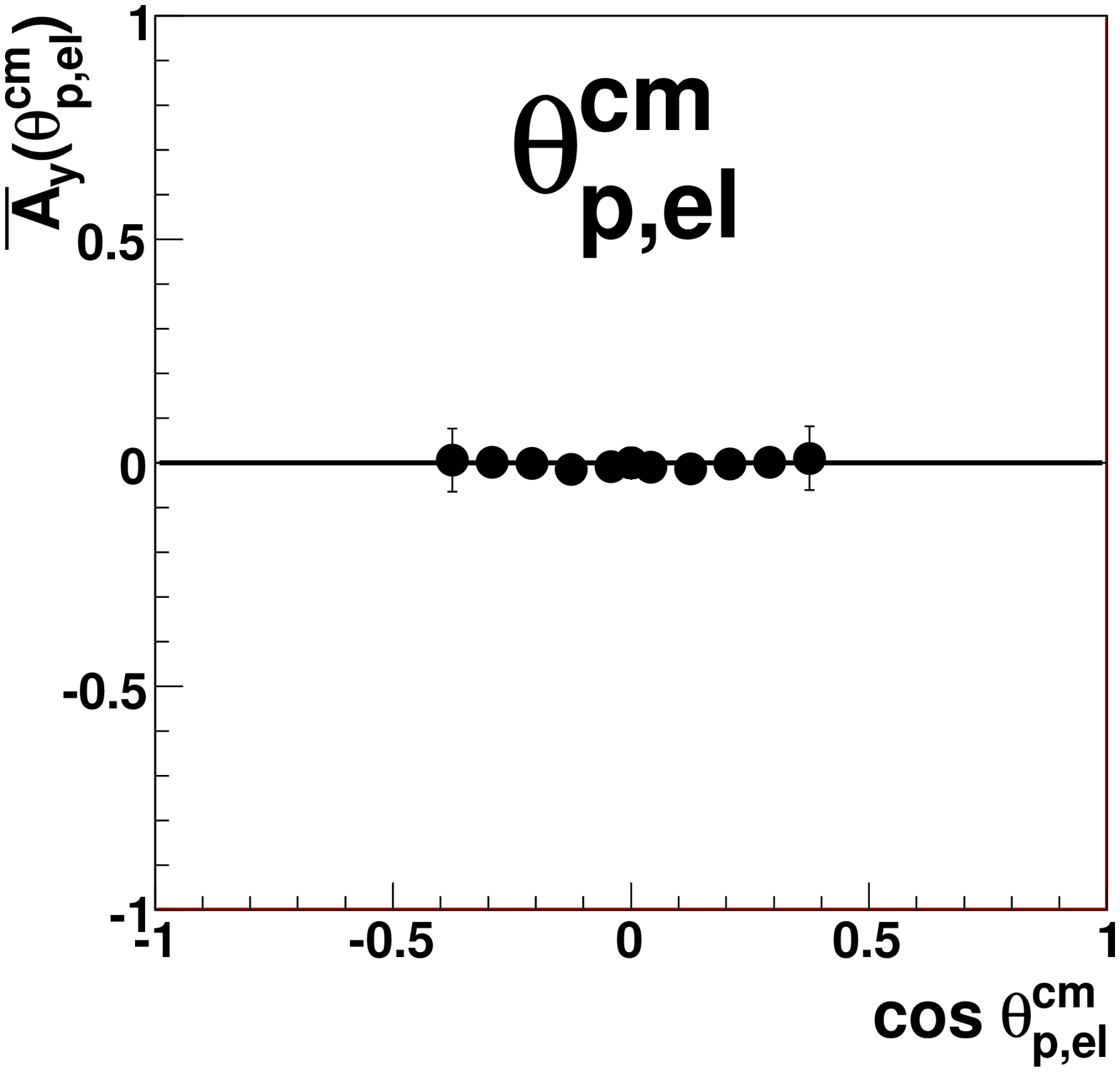}
\includegraphics[width=10pc]{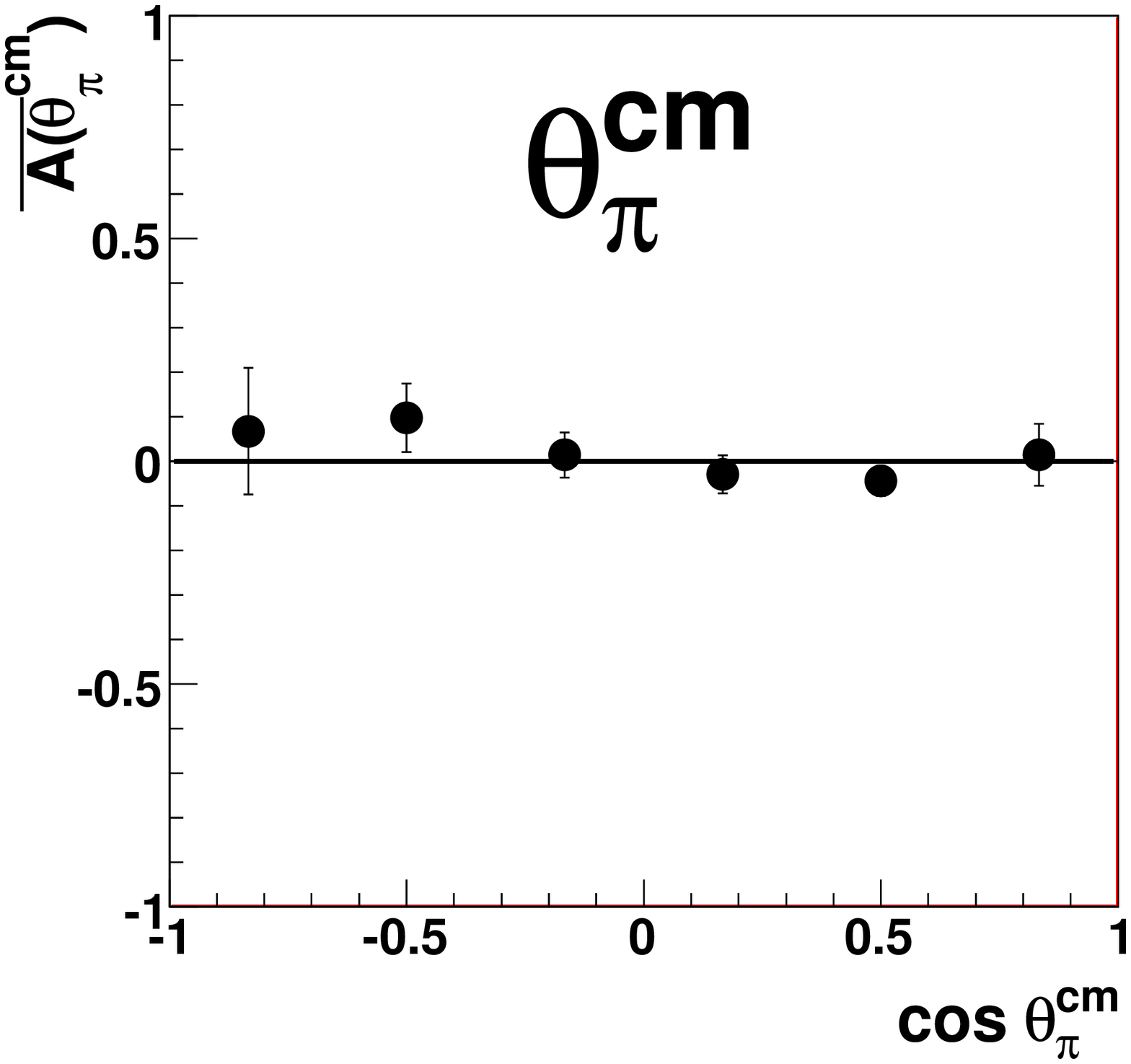}
\end{center}
\caption{"Anti-analyzing power" (instrumental asymmetry) as defined in eq.(5) for elastic scattering
  (left) and two-pion production events in dependence of the two-pion polar
  angle $\Theta_{\pi}^{cm}$(right).}
\end{figure}

\section{Results}
\label{sec:3}

\subsection{Cross Sections - Unpolarized}
\label{sec:3.1}

Due to the identity of the collision
partners in the entrance channel the angular distributions in the overall
center-of-mass system have to be symmetric about 90 $^\circ$, i.e. the
full information about the reaction channels is contained already in the
interval $0^\circ\leq \Theta^{cm}\leq 90^\circ$. Deviations from this
symmetry in the data indicate systematic uncertainties in the
measurements. Hence we plot  - where appropriate - the full angular range, in
order to show the absence of such type of systematic errors in our
measurement.  \\\\


Tab. 1: Total cross sections $\sigma_{tot}$ at $T_p~\approx~$750 and 800 MeV
for the 
reaction $pp\rightarrow pp\pi^+\pi^-$ evaluated in this work and compared to
previous measurements.

\vspace{0.5cm}
\begin{tabular}{llll} 
\hline

 & ~~~~~~~~~~~~~~~~~~& $\sigma_{tot}$ [$\mu$b]~~~~~~~~~~\\

 & $T_p$ [MeV] & this work & previous\\

\hline

& 750 & 1.6 (2) & 1.6 (3)$^{a)}$ \\

&  &  & 1.3  (3)$^{b)}$\\

 & 800 & 4.1 (4) & 3 (1)$^{c)}$\\ 

\hline

 \end{tabular}\\

 $^{a)}$ Ref. \cite{WB,JJ} ~~~  $^{b)}$ Ref. \cite{JP} ~~~ $^{c)}$
 Ref. \cite{cverna}\\

\vspace{1cm}


The evaluated total cross sections for the $pp\pi^+\pi^-$ channel are given in
Tab. 1 and shown in Fig. 9 together with previous results. The uncertainties
assigned  
are based on systematics for acceptance and efficiency corrections as obtained
by variation of MC simulations for the detector response, where we have varied
the MC input assuming either pure phase space or some reasonable models for
the reaction under consideration. For 750 (800) MeV we have collected about
1600 (6000) good events, i.e., statistical uncertainties are of minor
importance for the uncertainties in the absolute total cross section.\\

For $T_p$ = 750 MeV total and differential cross sections for this reaction
channel are very well known from previous experiments at CELSIUS 
\cite{WB,JPthesis}. Hence we use the analysis of our data at this energy
primarily as a check of the reliability of our measurement and data analysis.
In fact, the differential distributions obtained at CELSIUS and COSY-TOF are
compatible within statistical uncertainties \cite{kress,AE}. As an example we
compare in Fig. 10 both data sets for the differential distributions of the
invariant mass $M_{\pi\pi}$ and the opening angle $\delta_{\pi\pi}$ between
the two pions in the 
overall cms. We also note that the values for $\sigma_{tot}$
agree very well within uncertainties.

\begin{figure}
\begin{center}
\includegraphics[width=20pc]{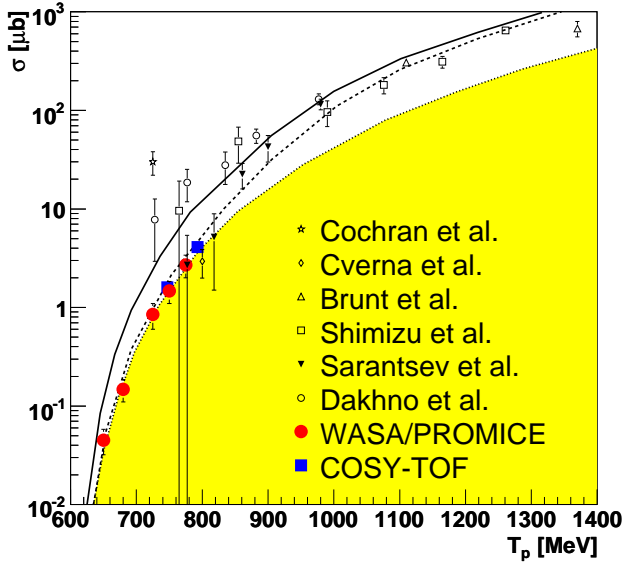}
\end{center}
\caption{Energy dependence of the total cross section. Shown are the data of
  this work (solid squares) together with those from CELSIUS \cite{WB,JP,JJ}
  (solid circles), Gatchina - old \cite{dakhno} (open circles) and new
  \cite{gat} (solid triangles), LAMPF \cite{cverna,coch} (open diamond and
  star, respectively), KEK \cite{shim} (open squares) and Berkeley \cite{brunt}
(open triangles). Solid and dashed curves correspond to theoretical
calculations of Ref.\cite{luis} with and without $pp$ FSI. The hatched area
represents the phasespace dependence adjusted arbitrarily  to the cross section
at $T_p$ = 750 MeV.  
 }
\end{figure}

\begin{figure}
\begin{center}
\includegraphics[width=10pc]{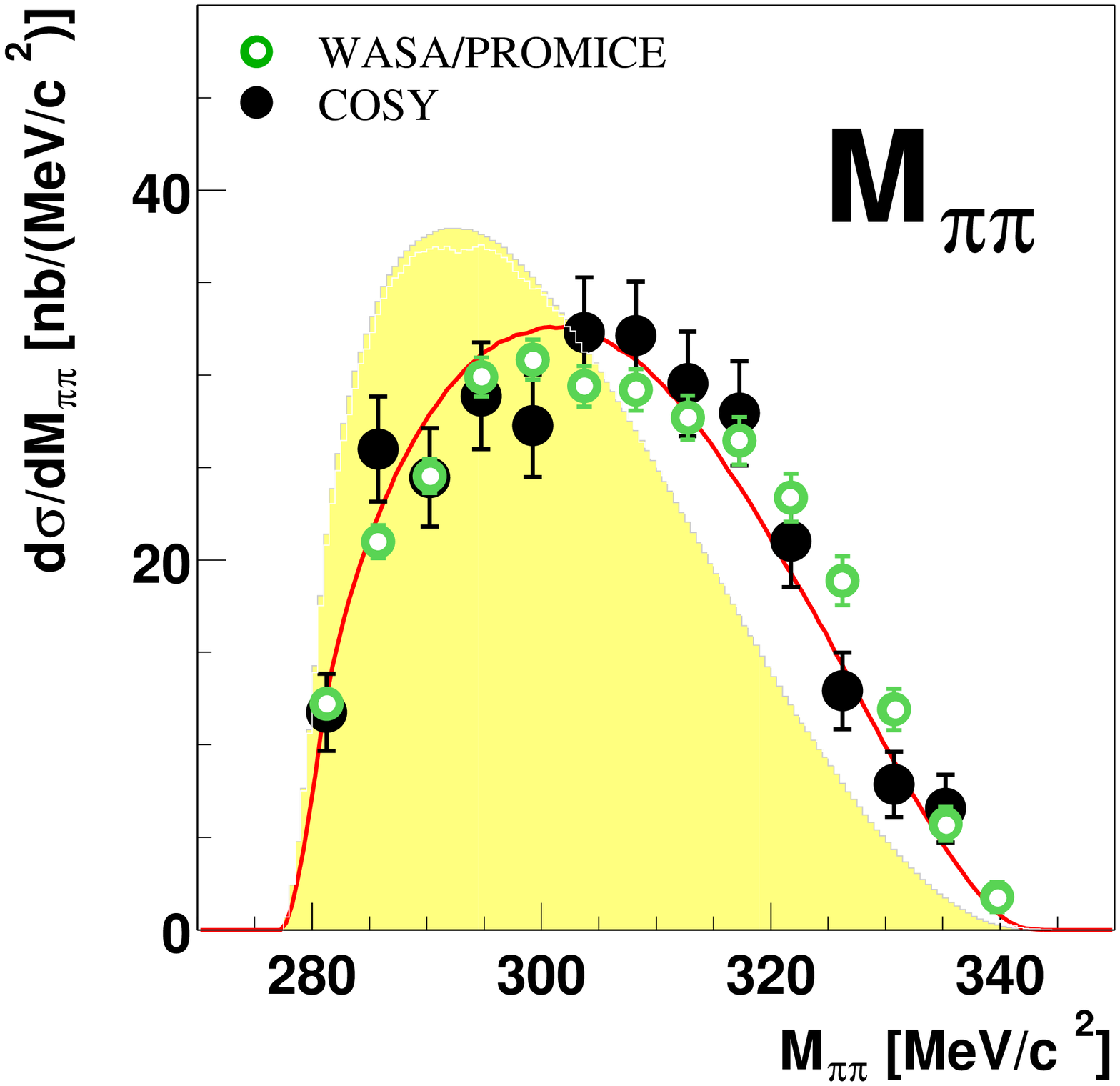}
\includegraphics[width=10pc]{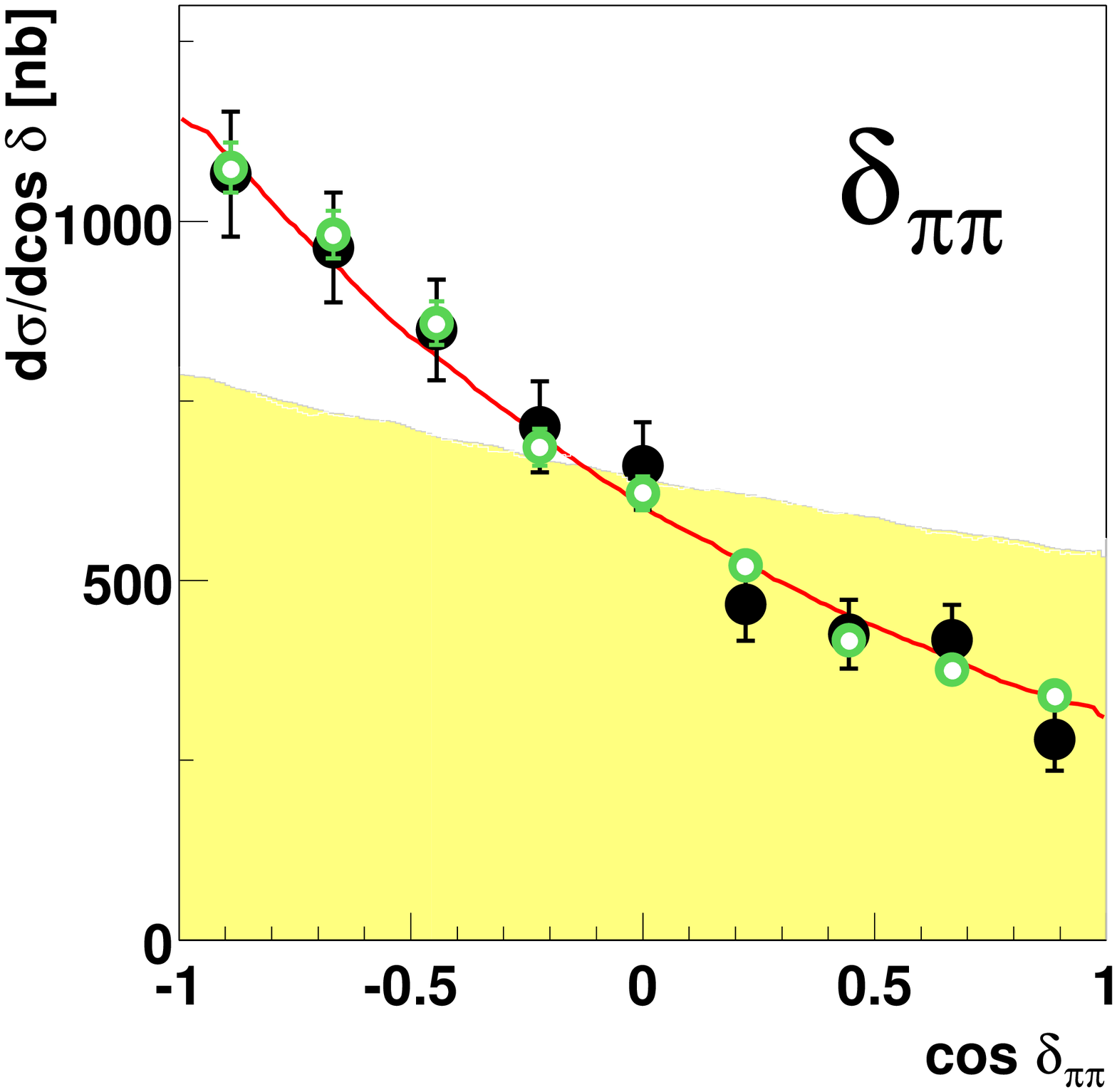}
\end{center}
\caption{Differential cross sections for the distributions of $\pi\pi$
  invariant mass $M_{\pi\pi}$ (left) and  $\pi\pi$ opening angle in the
  overall cms  $\delta_{\pi\pi}$ (right) at $T_p$ = 750 MeV. The data of this
  work (filled symbols) 
  are compared to those obtained at CELSIUS (open symbols) as well as to phase
  space distributions (shaded areas) and calculations according to
  eq.(6) (solid lines). CELSIUS data and theoretical curves are renormalized in area
  ($\sigma_{tot}$) to the data of this work.
} 
\end{figure}

\begin{figure}
\begin{center}
\includegraphics[width=10pc]{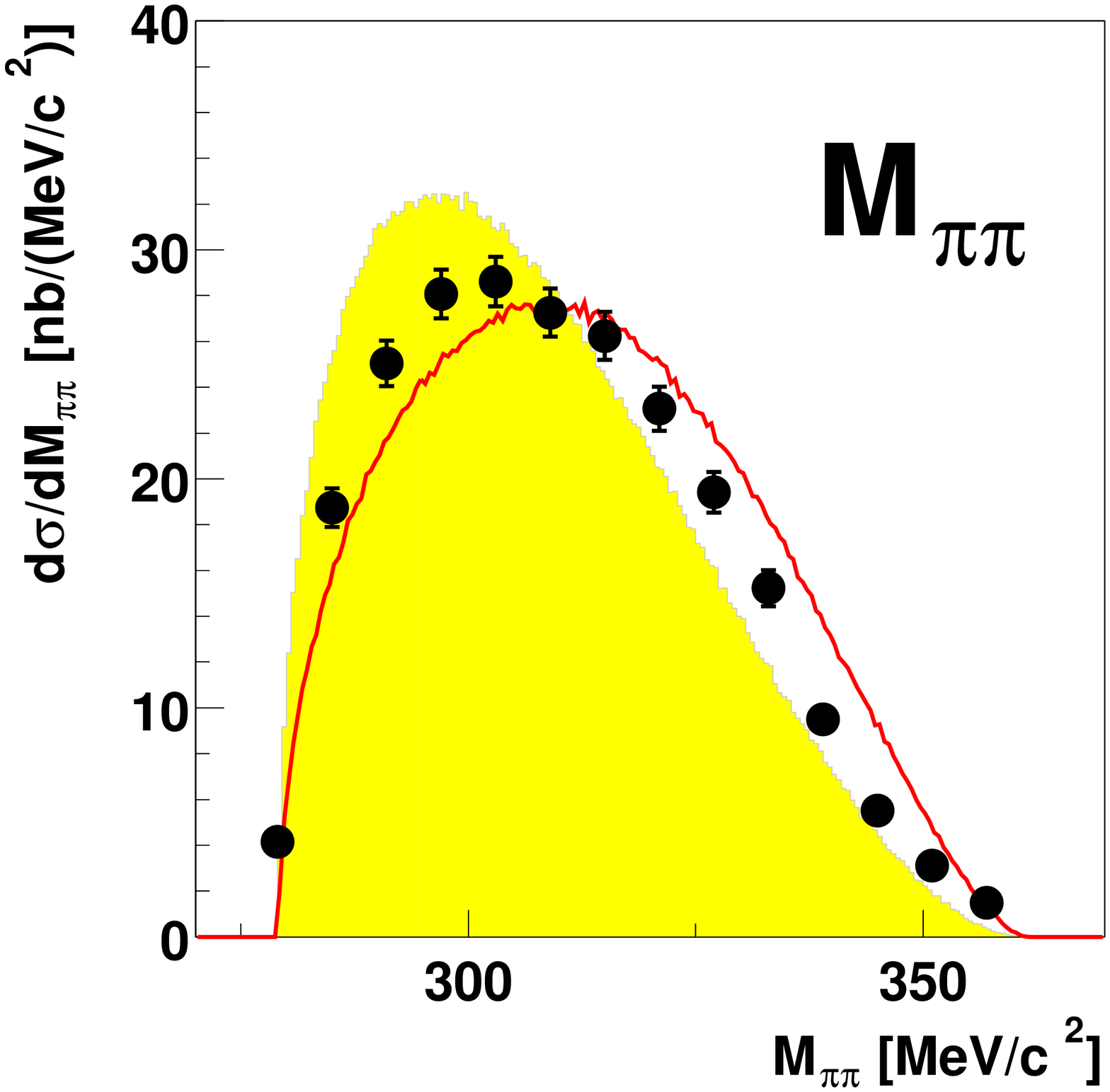}
\includegraphics[width=10pc]{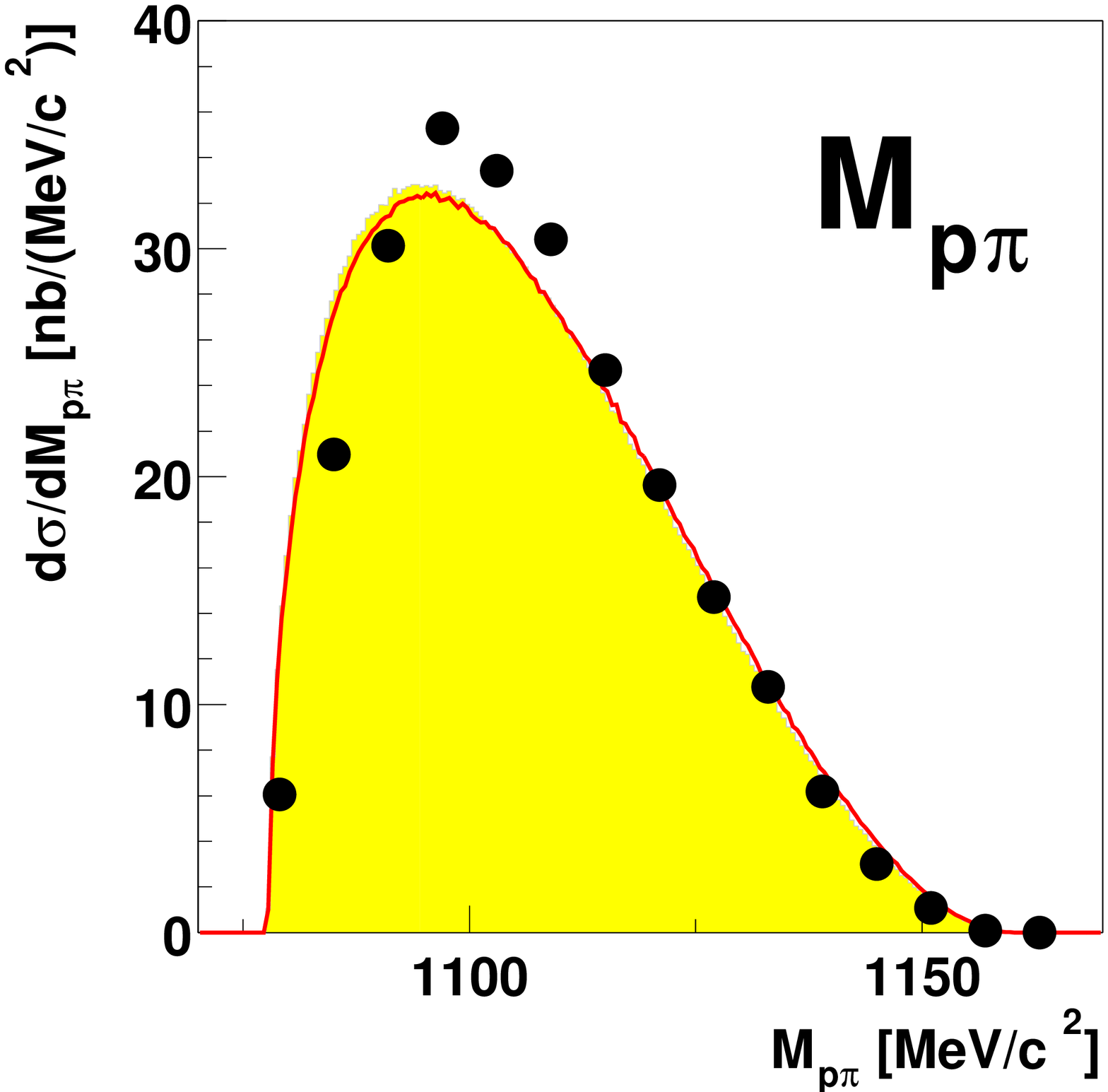}
\includegraphics[width=10pc]{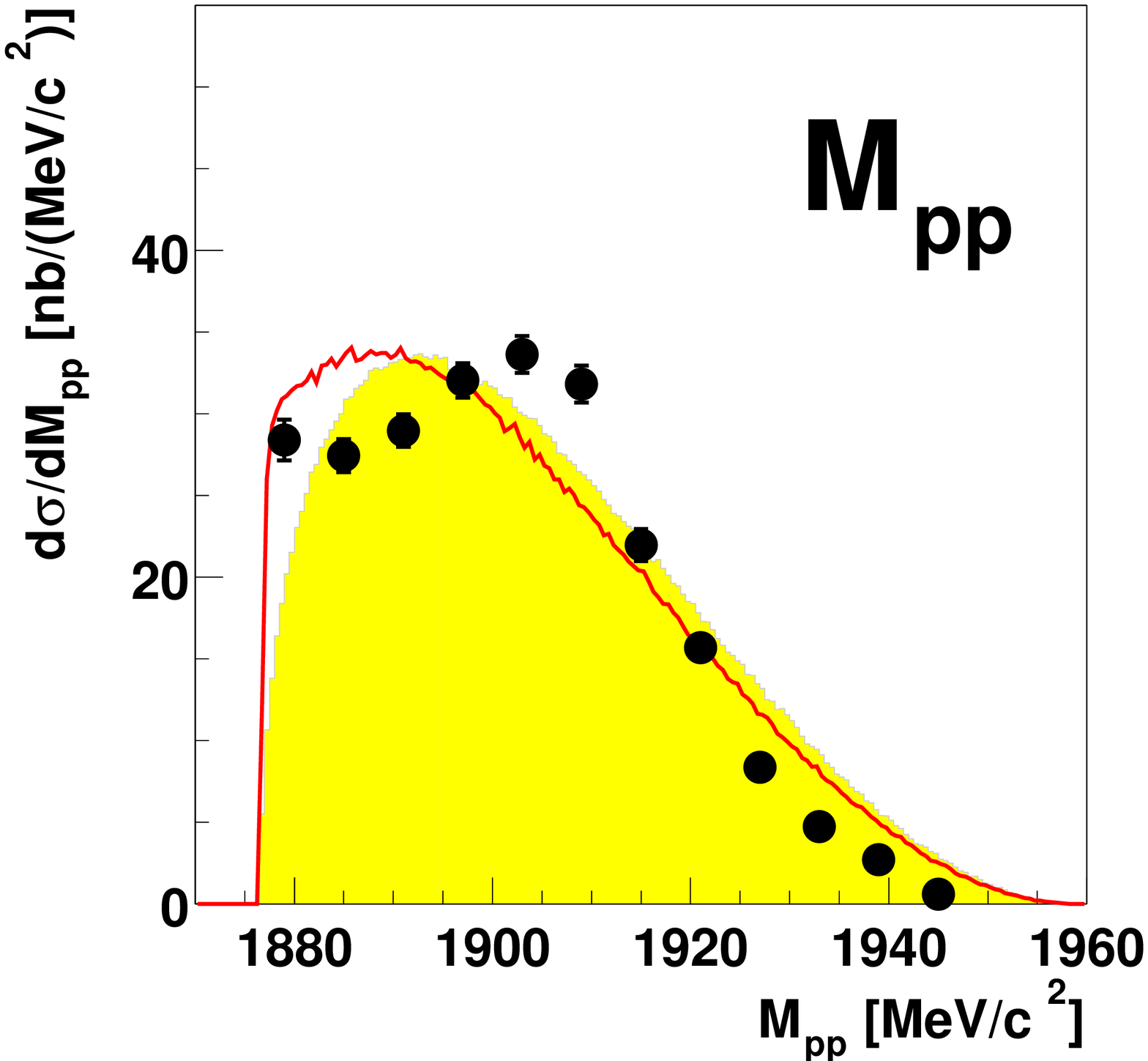}
\includegraphics[width=10pc]{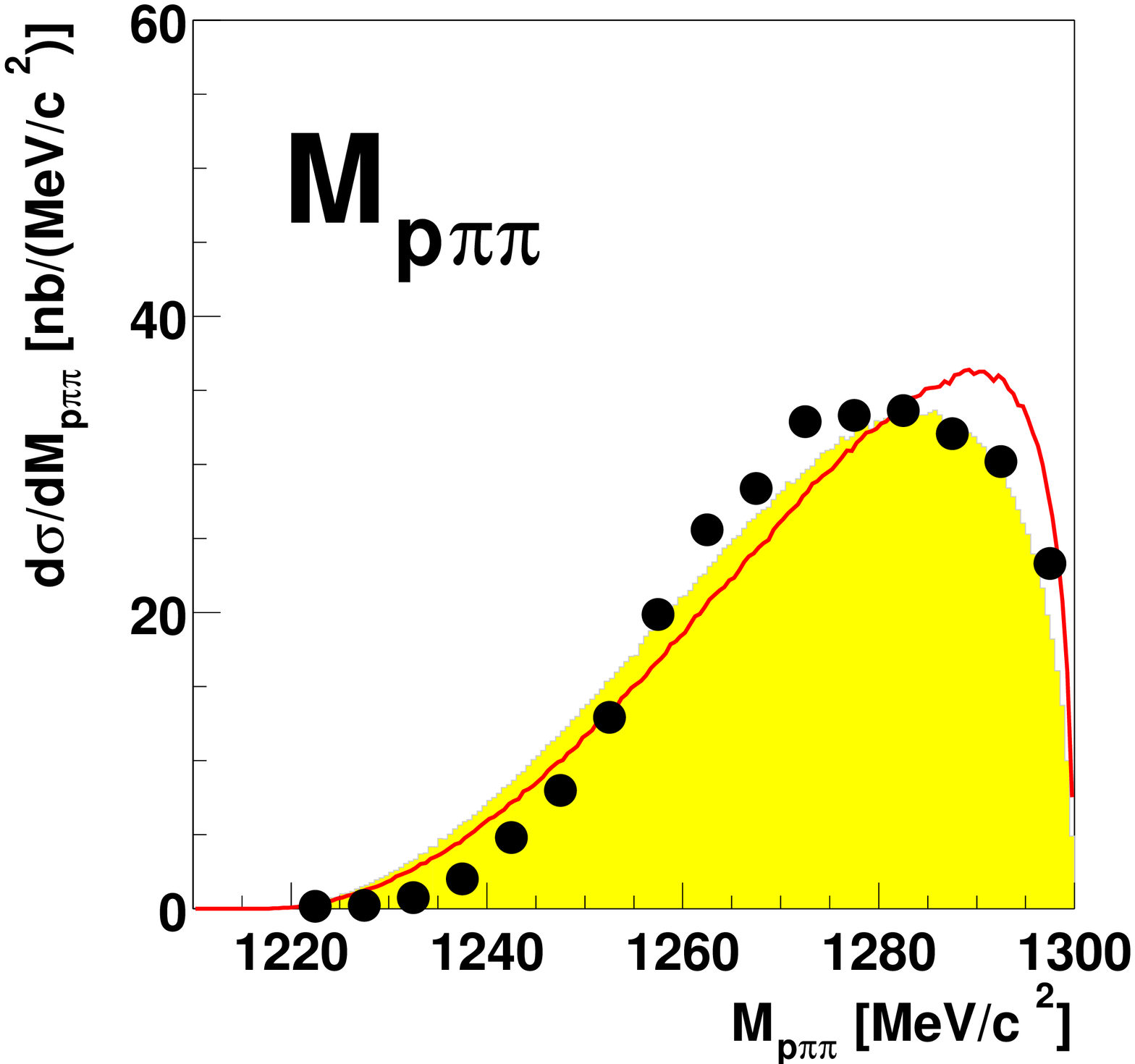}
\end{center}
\caption{Results of this work for differential distributions of the invariant
  masses $M_{pp}$,$M_{\pi\pi}$,$M_{p\pi}$ and $M_{p\pi\pi}$ at $T_p$ = 800
  MeV. They are compared to phasespace distribuions (shaded areas) as well as
  to calculations according to  eq.(6) (solid lines).}
\end{figure}

\begin{figure}
\begin{center}
\includegraphics[width=10pc]{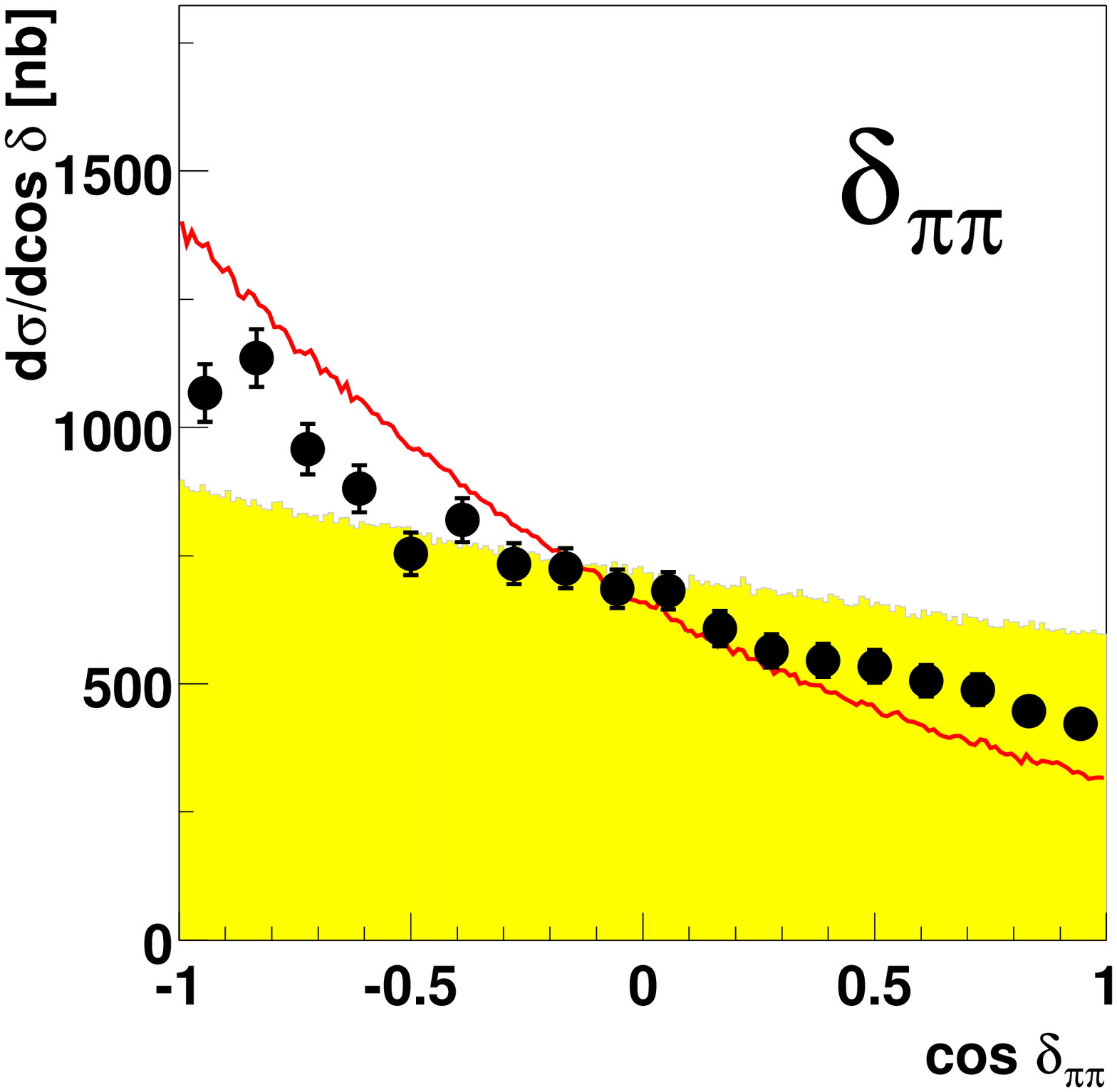}
\includegraphics[width=10pc]{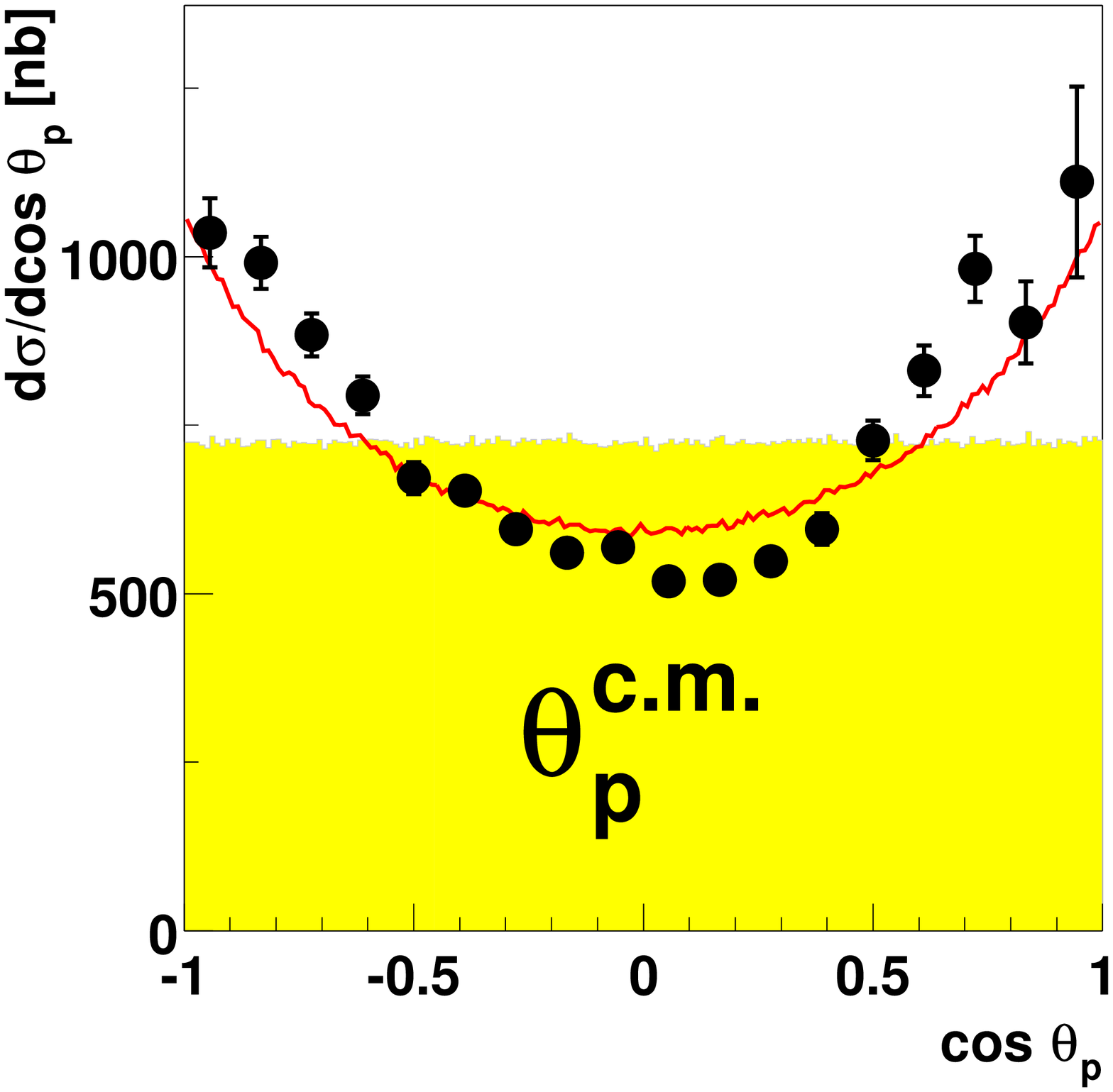}
\includegraphics[width=10pc]{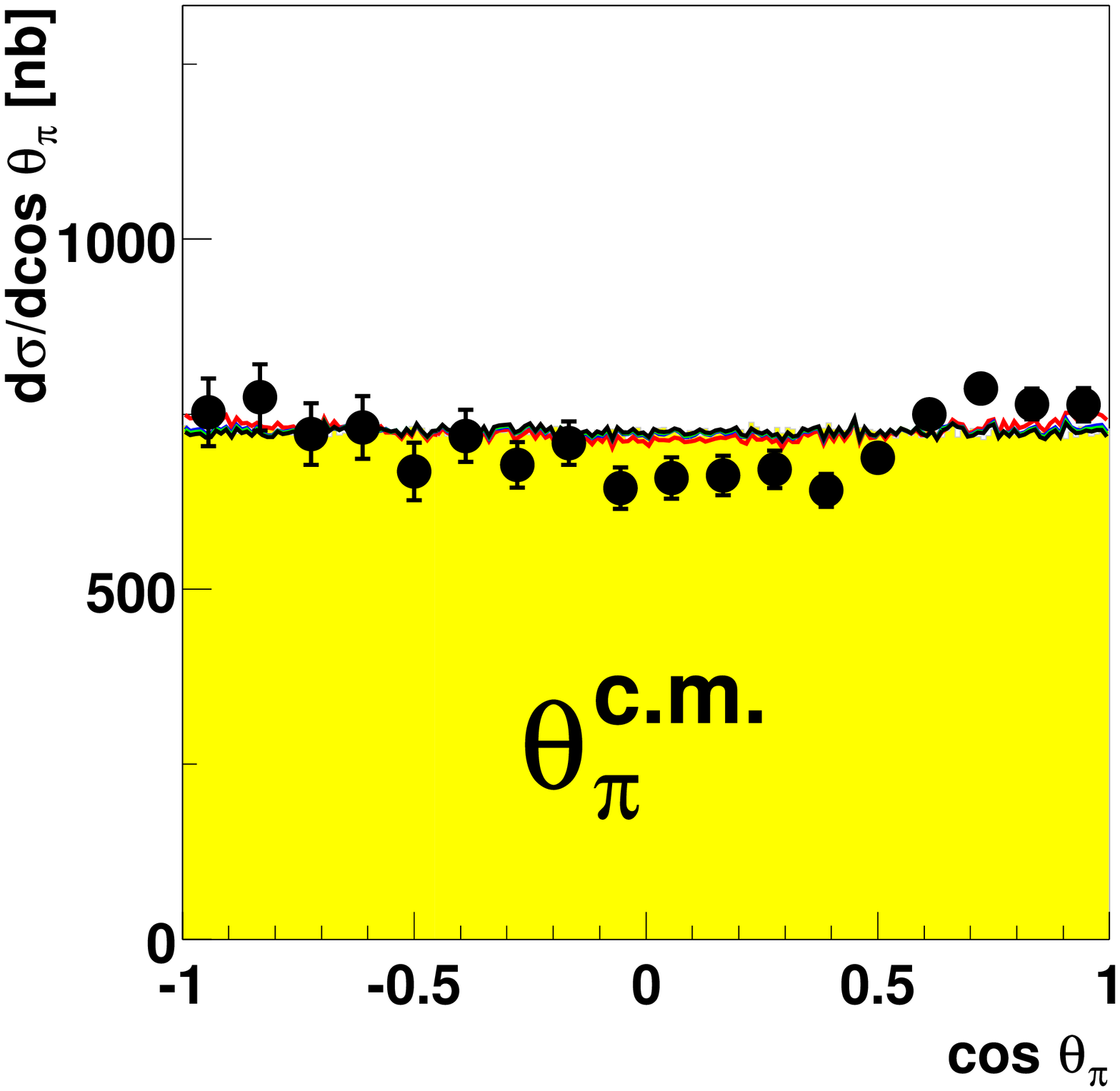}
\includegraphics[width=10pc]{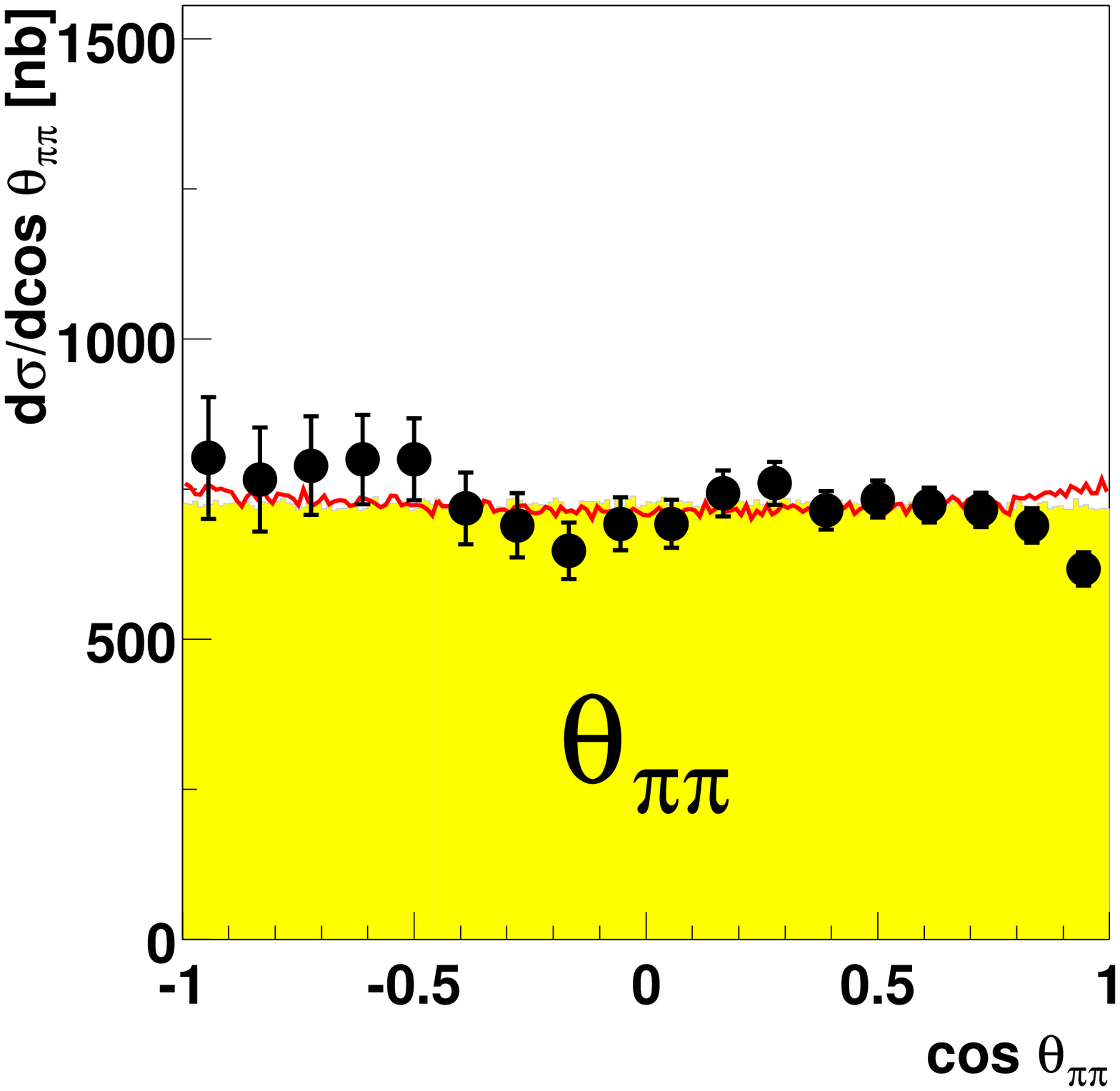}
\end{center}
\caption{The same as Fig. 11, but for  of the opening angle between the two
  pions $\delta_{\pi\pi}$, the polar angles of protons $\Theta_{p}^{cm}$,
  pions $\Theta_{\pi}^{cm}$ and of the two-pion system  $\Theta_{\pi\pi}^{cm}$
  -  all in the overall cms.}
\end{figure}

\begin{figure}
\begin{center}
\includegraphics[width=10pc]{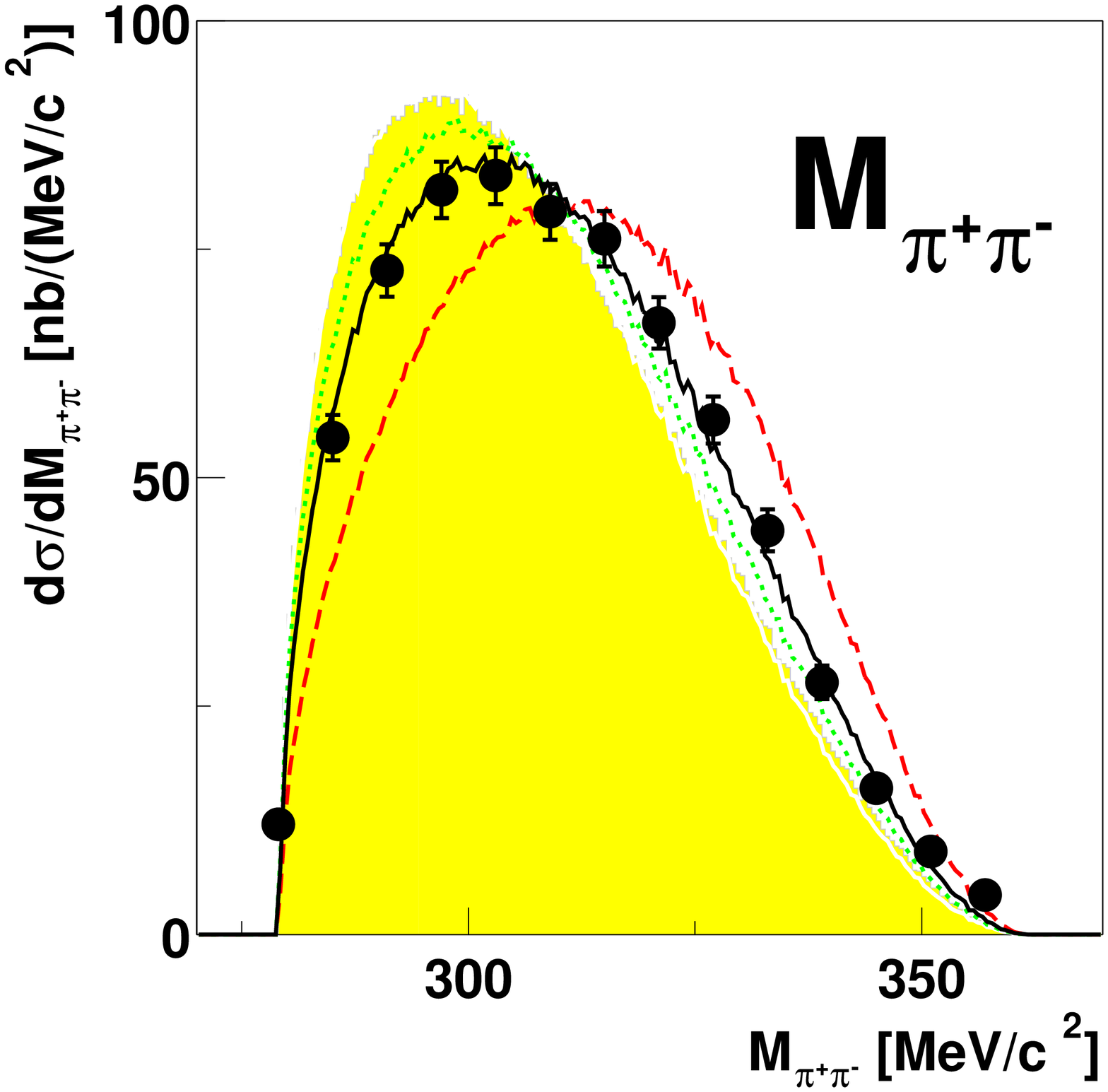}
\includegraphics[width=10pc]{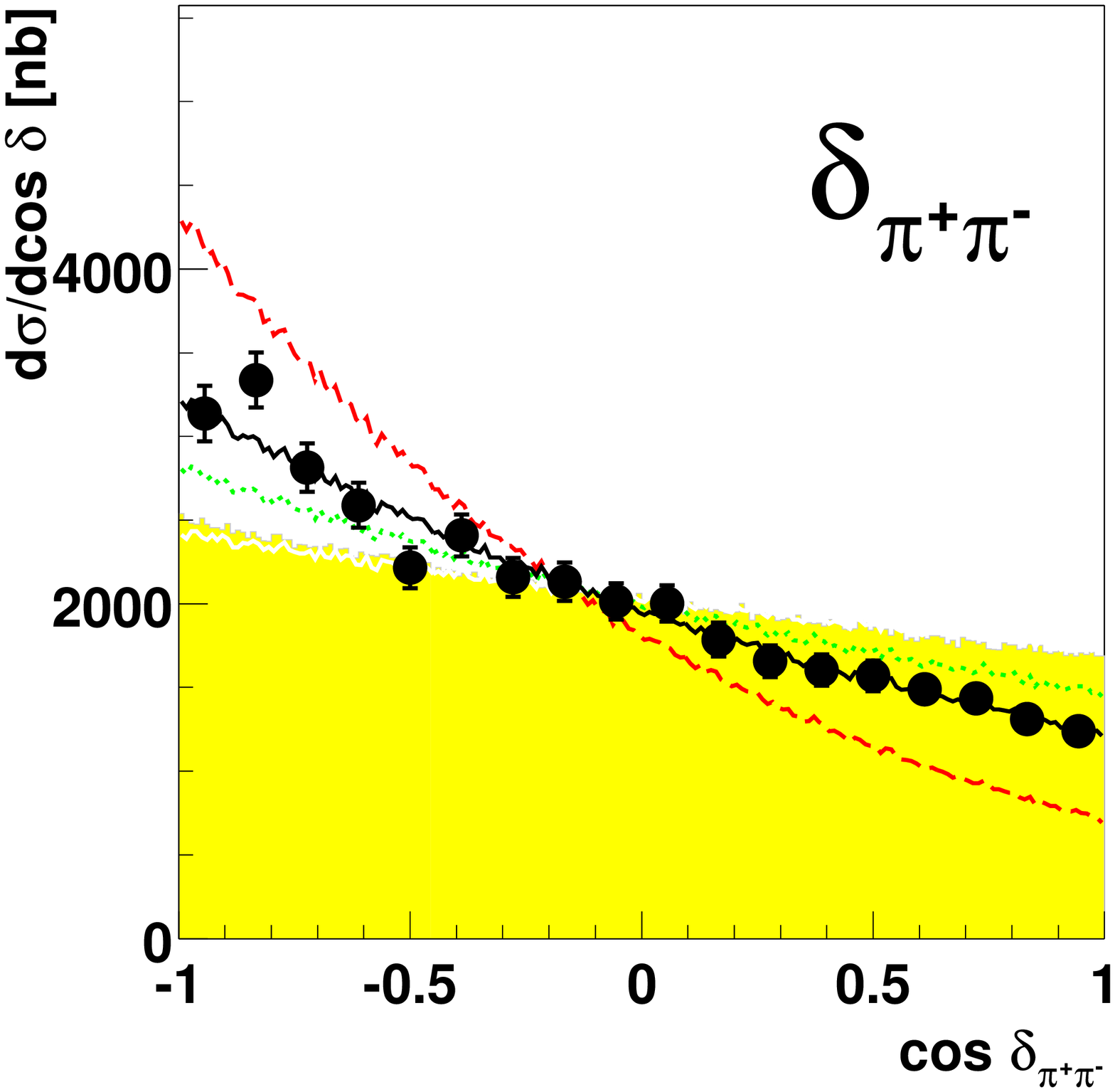}
\end{center}
\caption{Differential cross sections for the distributions of $\pi\pi$
  invariant mass $M_{\pi\pi}$ (left) and  $\pi\pi$ opening angle in the
  overall cms  $\delta_{\pi\pi}$ (right) at $T_p$ = 800 MeV. The data of this
  work are compared  to phase
  space distributions (shaded areas) and calculations according to
  Ref.\cite{luis}. The dashed lines show the original calculations of
  Ref. \cite{luis} renormalized to the data in area, the 
  solid lines the result, when we reduce the amplitude for the branch $N^* \to
  \Delta \pi$ by a factor of two relative to the branch $N^* \to N\sigma$.
}
\end{figure}

In contrast to $T_p=750$ MeV there exists merely a
single-arm measurement for $T_p$ = 800 MeV, namely the one performed at LAMPF
\cite{cverna} with an estimated total cross section of 3 $\mu$b. 

Differential cross sections for invariant masses and
angular distributions obtained from our measurements are shown in Figs. 10 -
13. 

\subsection{Analyzing Powers}
\label{sec:3.2}

These measurements are the first ones with polarized beam, which  have
been conducted for the two-pion production channel in $NN$ collisions. The
angular distributions for the analyzing power have been obtained as described
in detail in section 2.

For $T_p$ = 750 MeV, where the statistics was quite moderate, we show in
Fig. 13 our results for the $A_y$ distributions on the angle of any proton
$\Theta_p^{cm}$, of any pion $\Theta_{\pi}^{cm}$, of the $\pi\pi$ system
$\Theta_{\pi\pi}^{cm}$ and of any $p\pi$ system $\Theta_{p\pi}^{cm}$ - all in
the overall cms.

For $T_p$ = 800 MeV, where we have much better statistics, we show in addition
to these distributions (Fig. 15) also the ones, where we differentiated
between $\pi^+$ and $\pi^-$ particles with help of the delayed pulse technique
 (Fig. 16). Finally we display in Fig. 17 the analyzing powers for the angles
 $\Theta_{\pi}^{\pi\pi}$ of any pion in the $\pi\pi$ subsystem and for the
 angles $\Theta_p^{pp}$ of any proton in the $pp$ subsystem, both taken in the
 Jackson frame, i.e., having the beam axis as z-axis.

\begin{figure}
\begin{center}
\includegraphics[width=10pc]{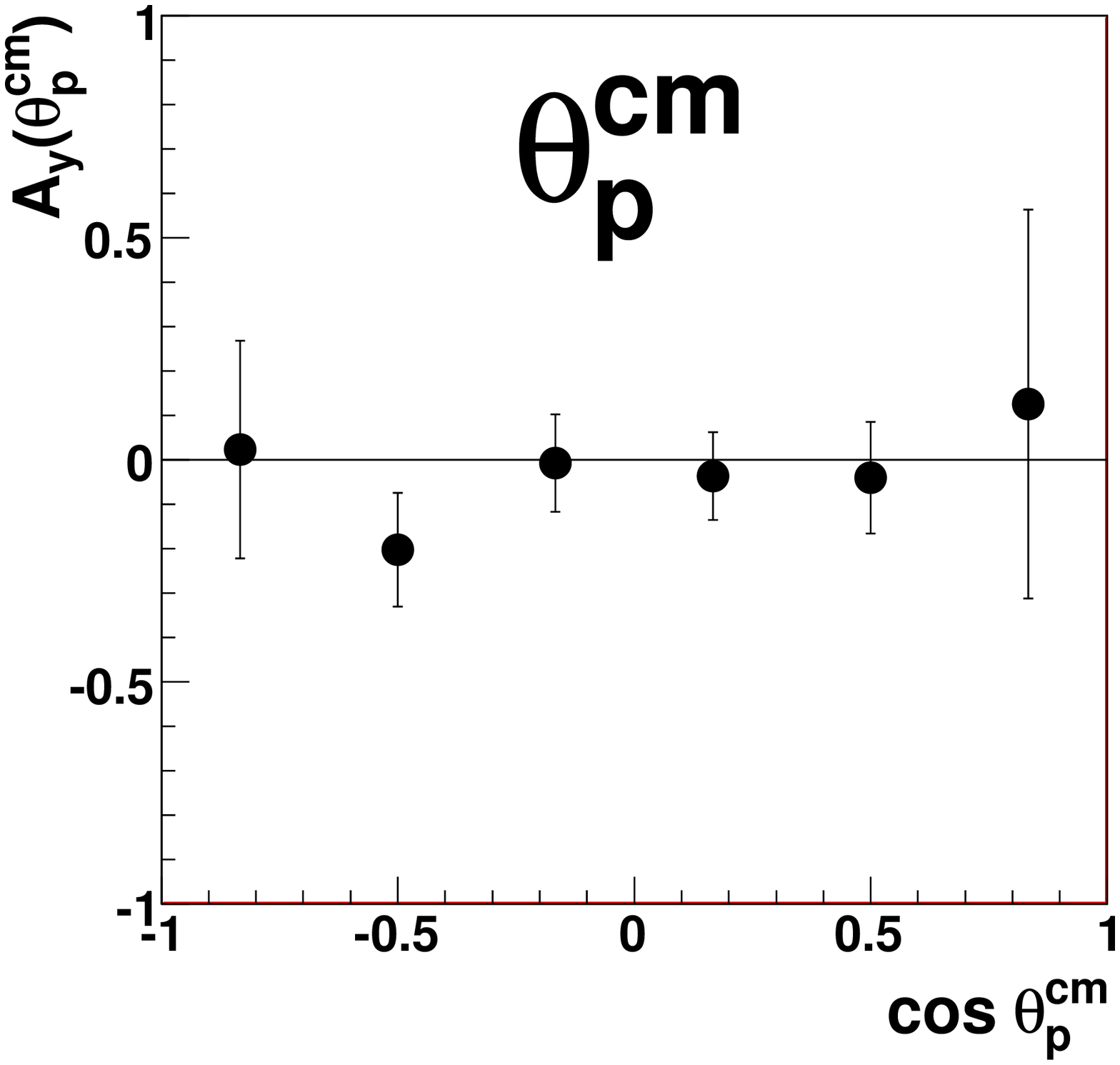}
\includegraphics[width=10pc]{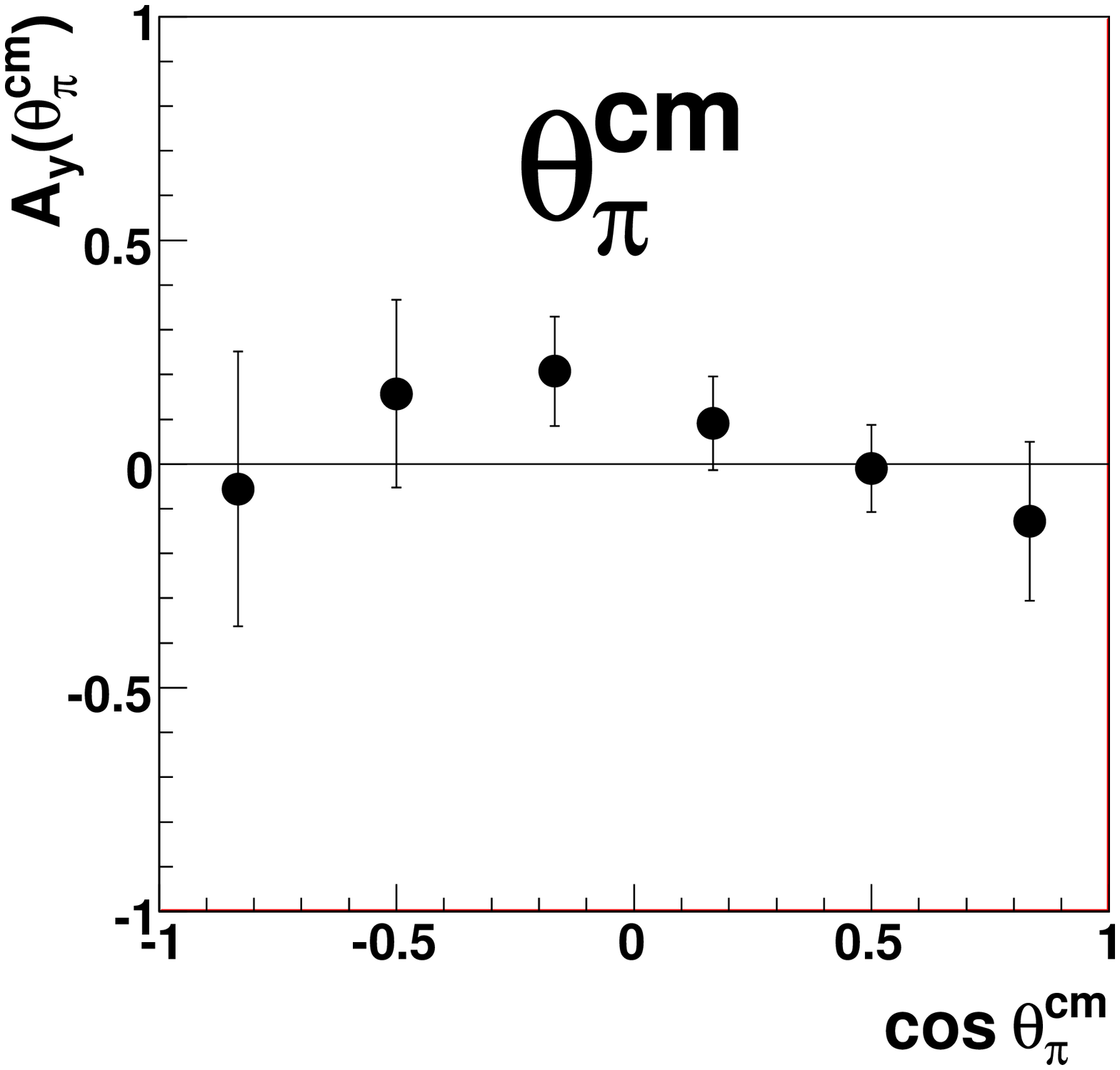}
\includegraphics[width=10pc]{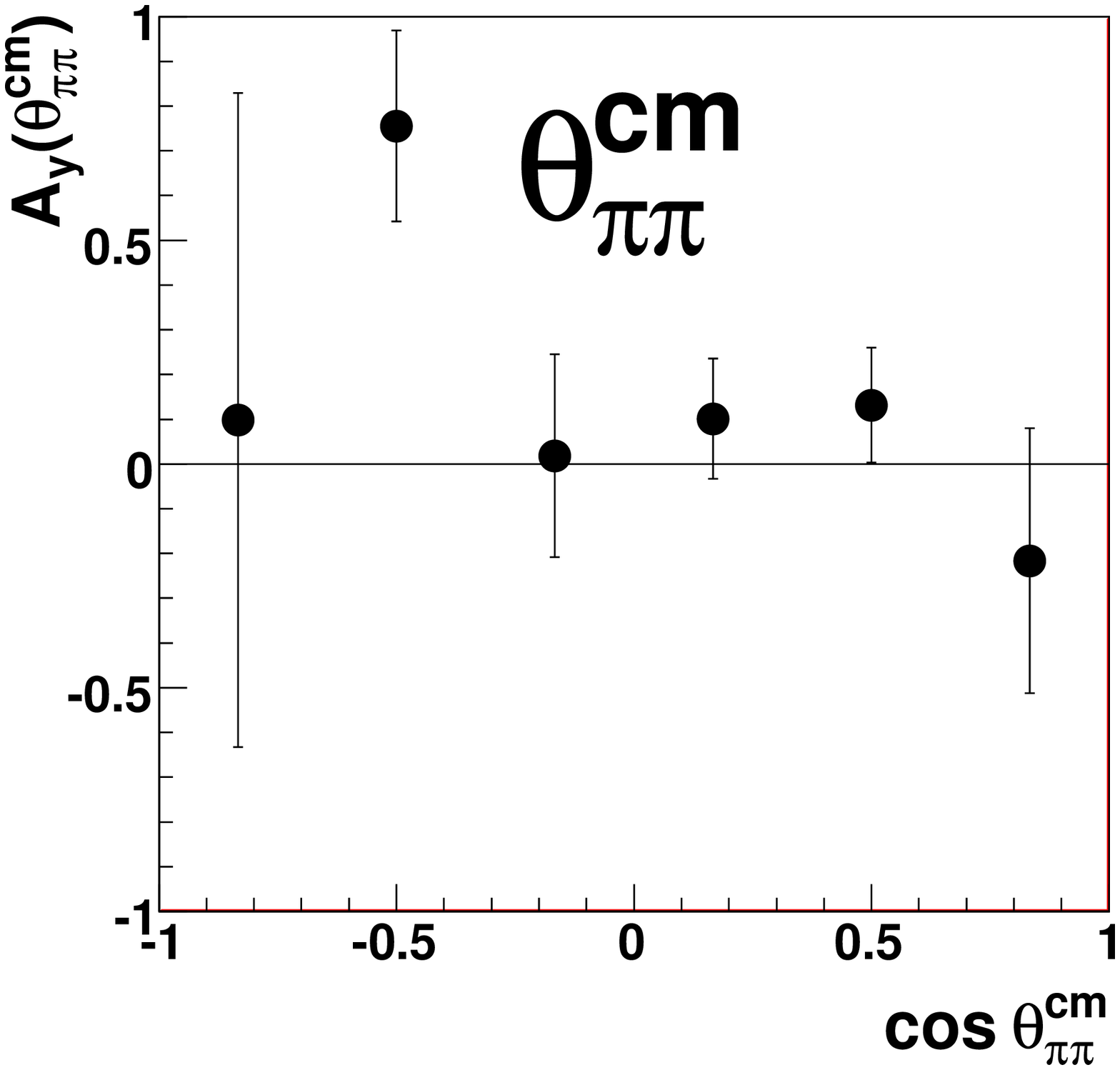}
\includegraphics[width=10pc]{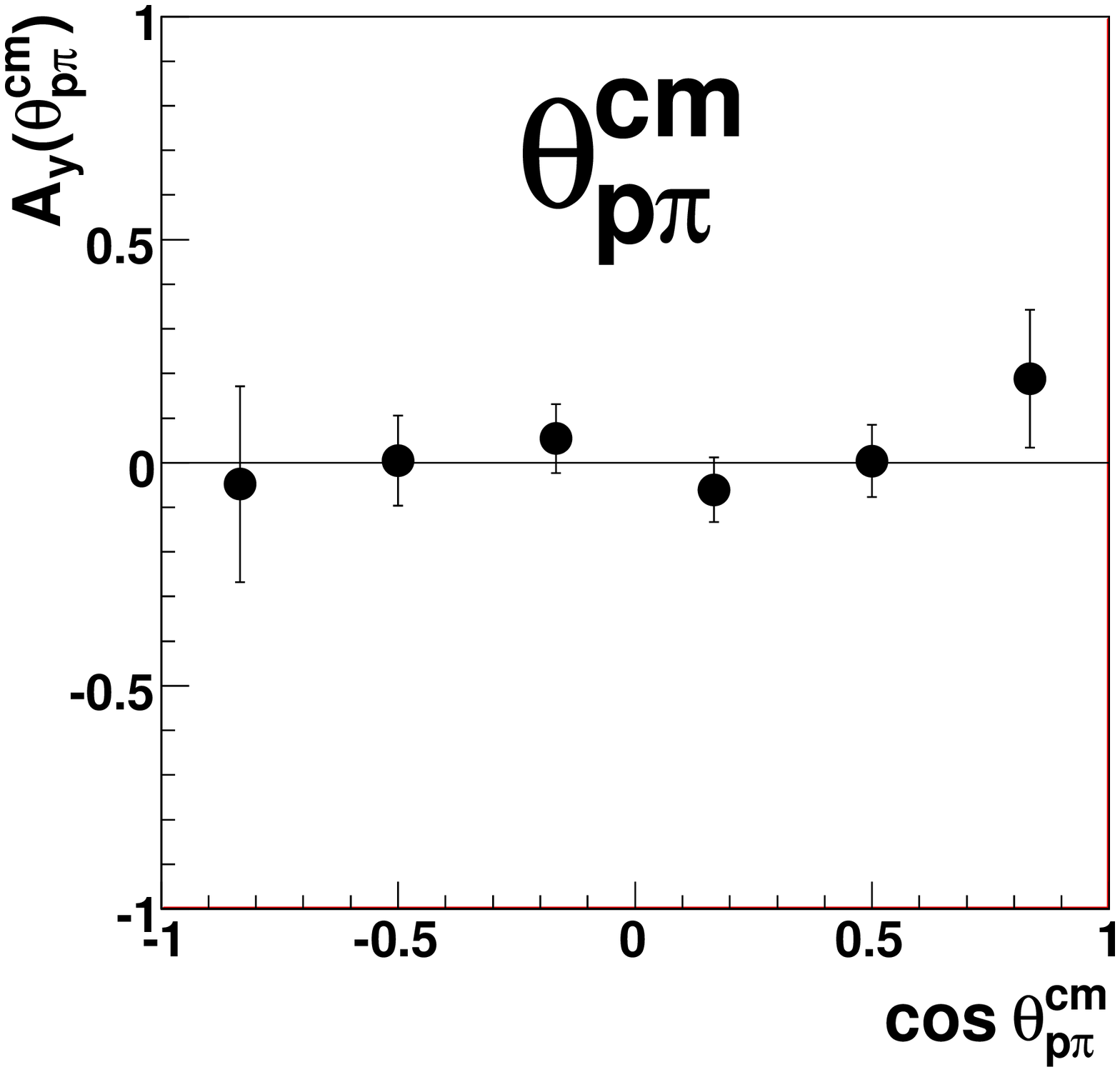}

\end{center}
\caption{Analyzing power distributions for the polar angles $\Theta_{p}^{cm}$,
  $\Theta_{\pi}^{cm}$, $\Theta_{\pi\pi}^{cm}$ and $\Theta_{p\pi}^{cm}$ at
  $T_p$ = 750 MeV.  }
\end{figure}

\begin{figure}
\begin{center}
\includegraphics[width=10pc]{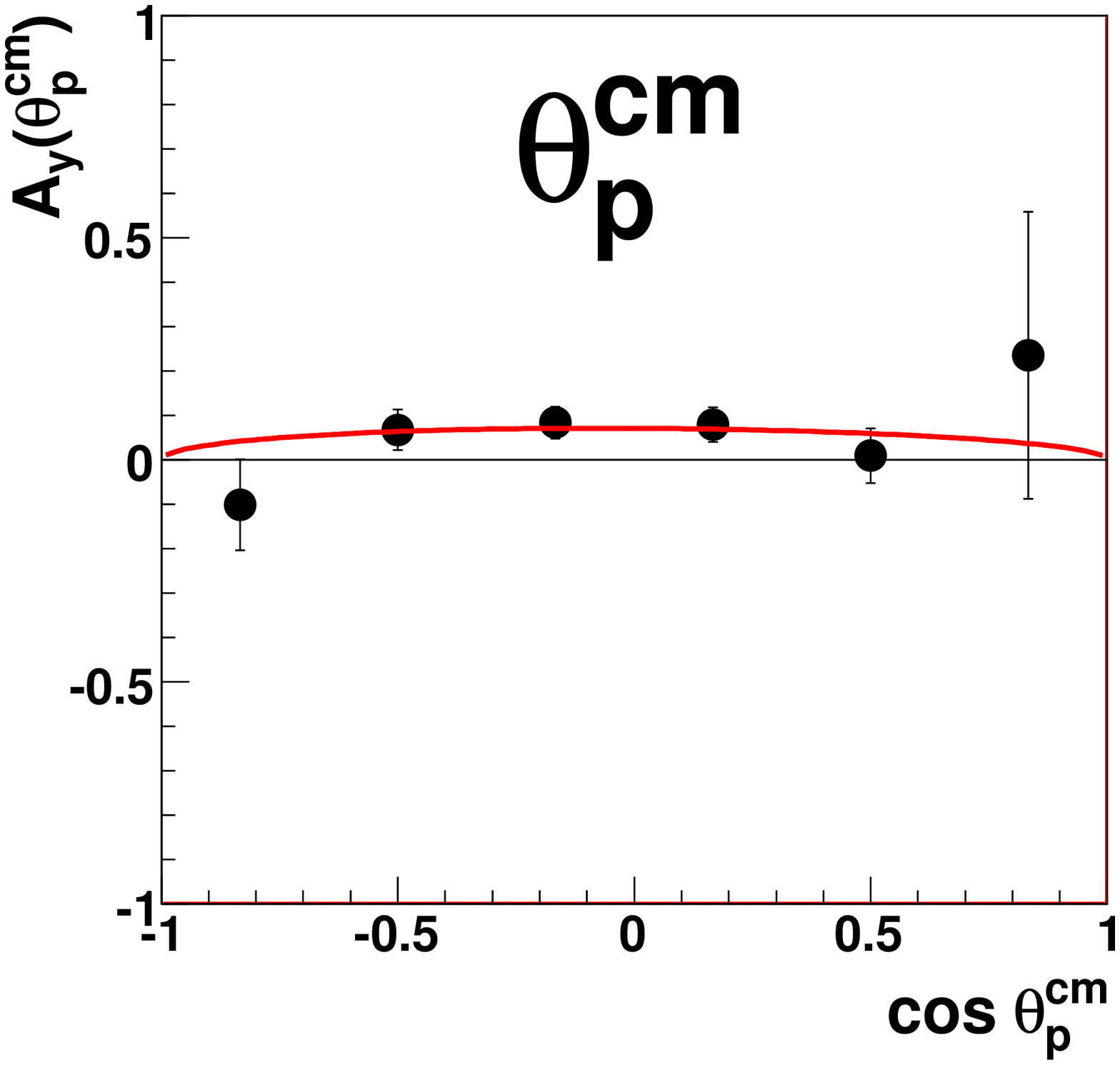}
\includegraphics[width=10pc]{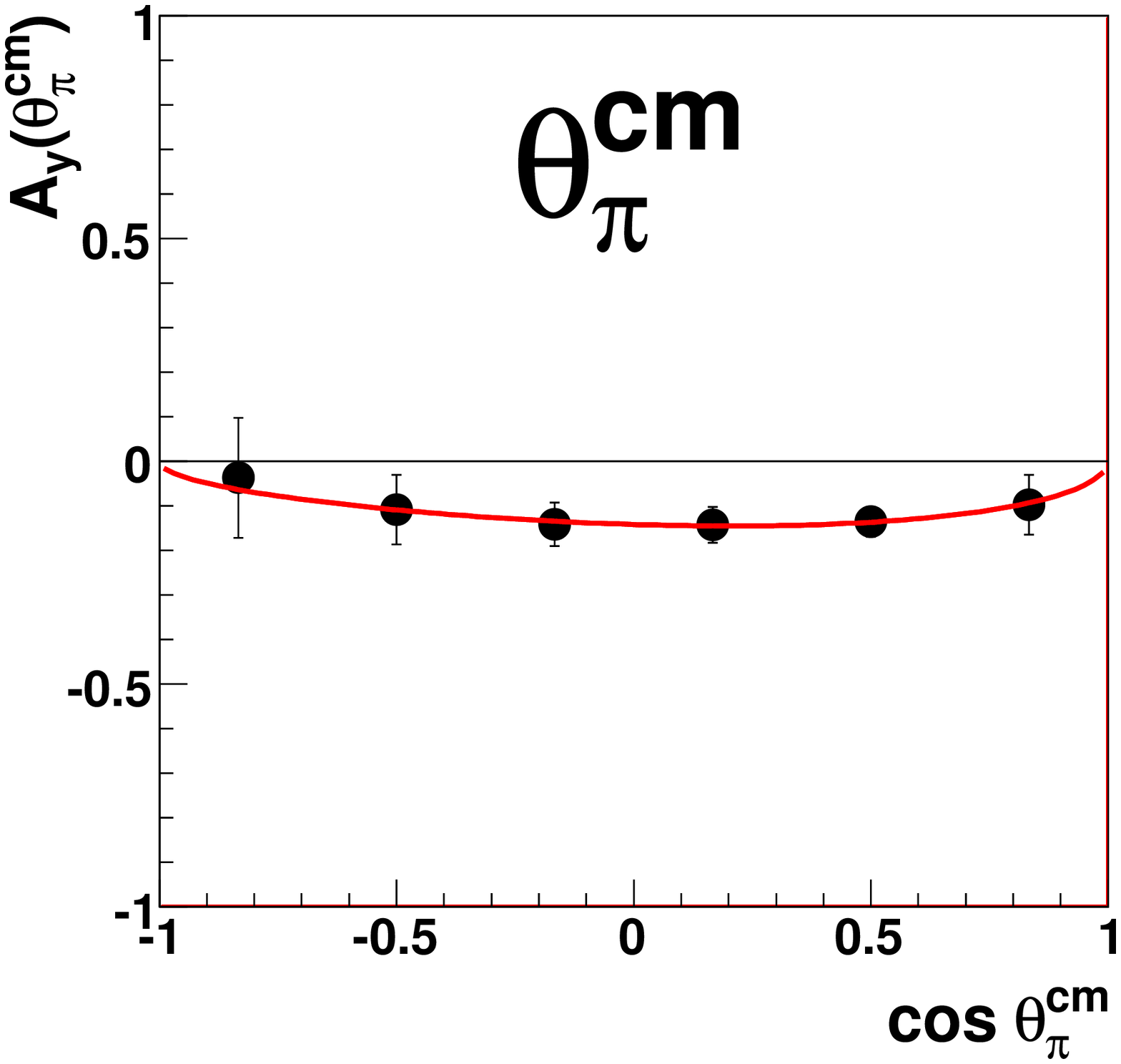}
\includegraphics[width=10pc]{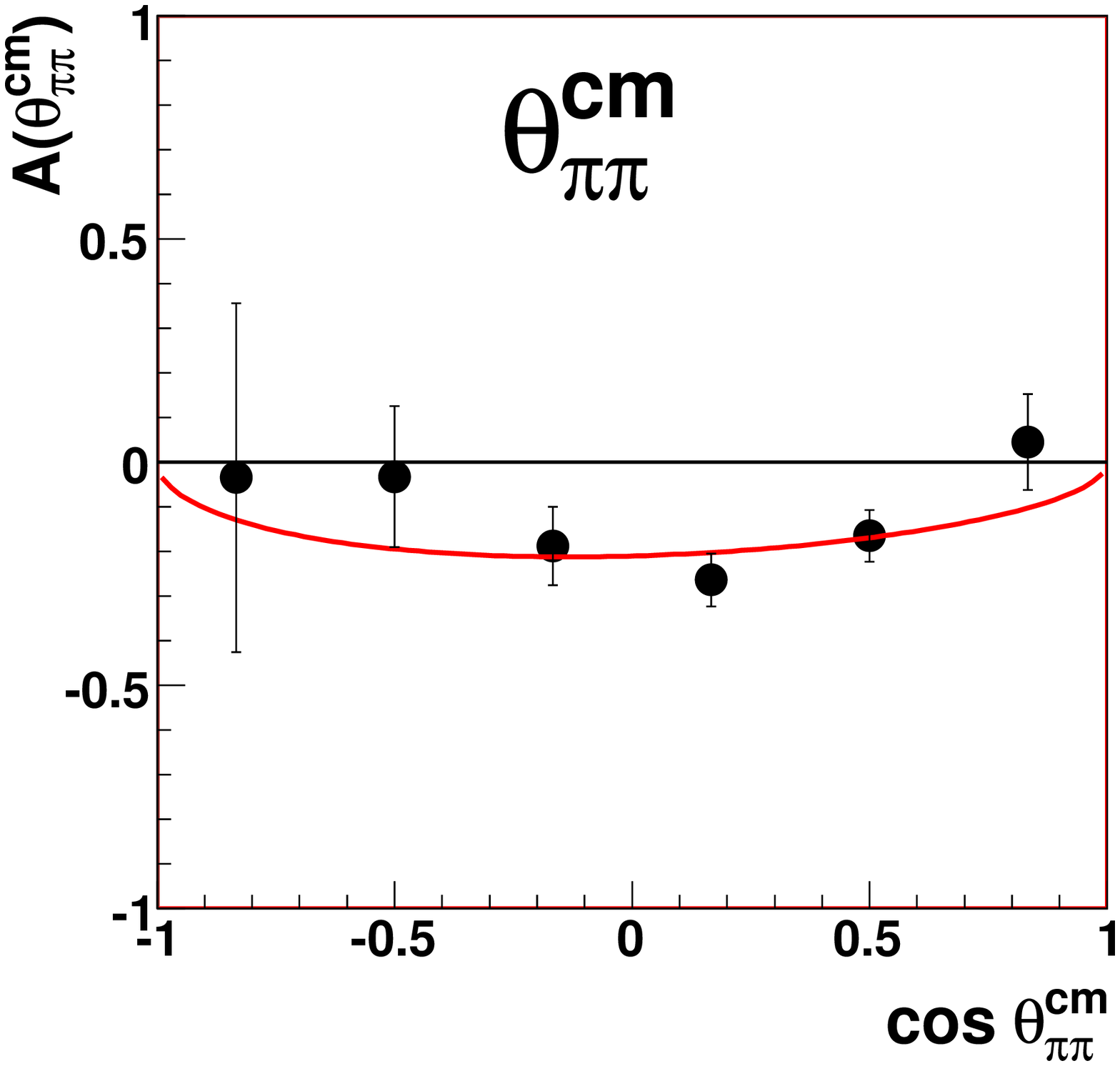}
\includegraphics[width=10pc]{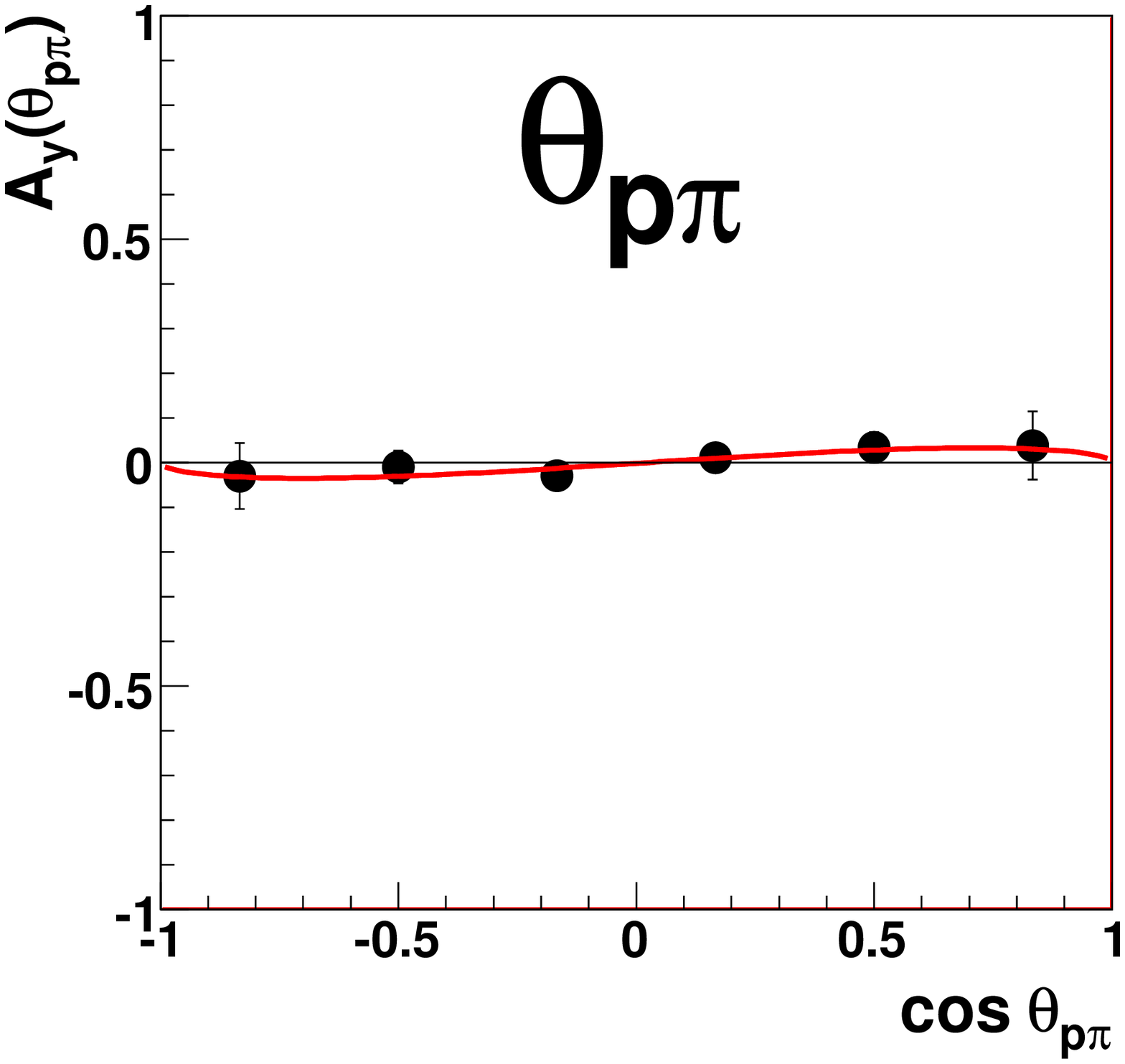}
\end{center}
\caption{Analyzing power distributions for the polar angles $\Theta_{p}^{cm}$,
  $\Theta_{\pi}^{cm}$, $\Theta_{\pi\pi}^{cm}$ and $\Theta_{p\pi}^{cm}$ at
  $T_p$ = 800 MeV. The solid curves show a fit according to eq.(7).  }
\end{figure}

\begin{figure}
\begin{center}
\includegraphics[width=10pc]{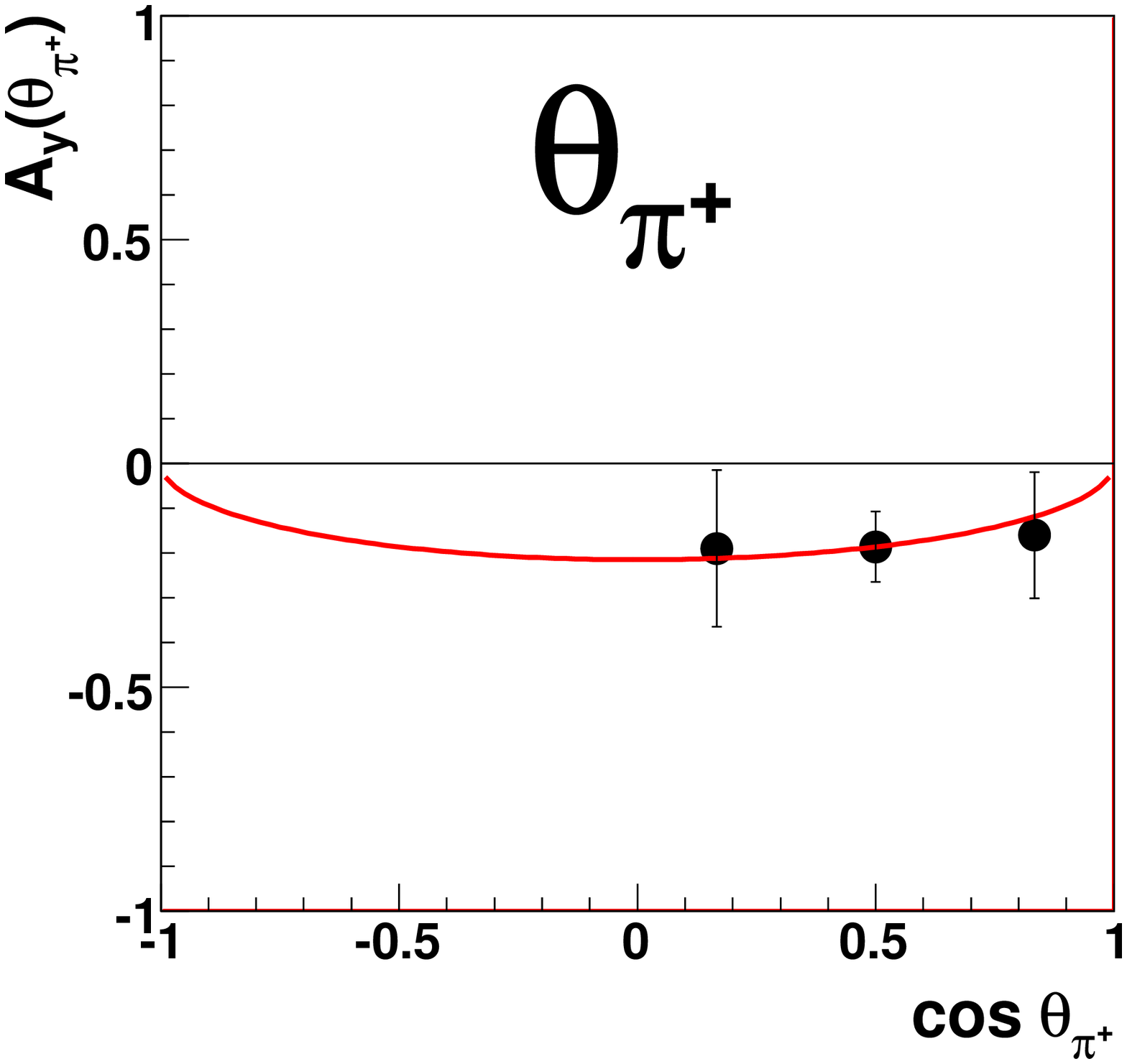}
\includegraphics[width=10pc]{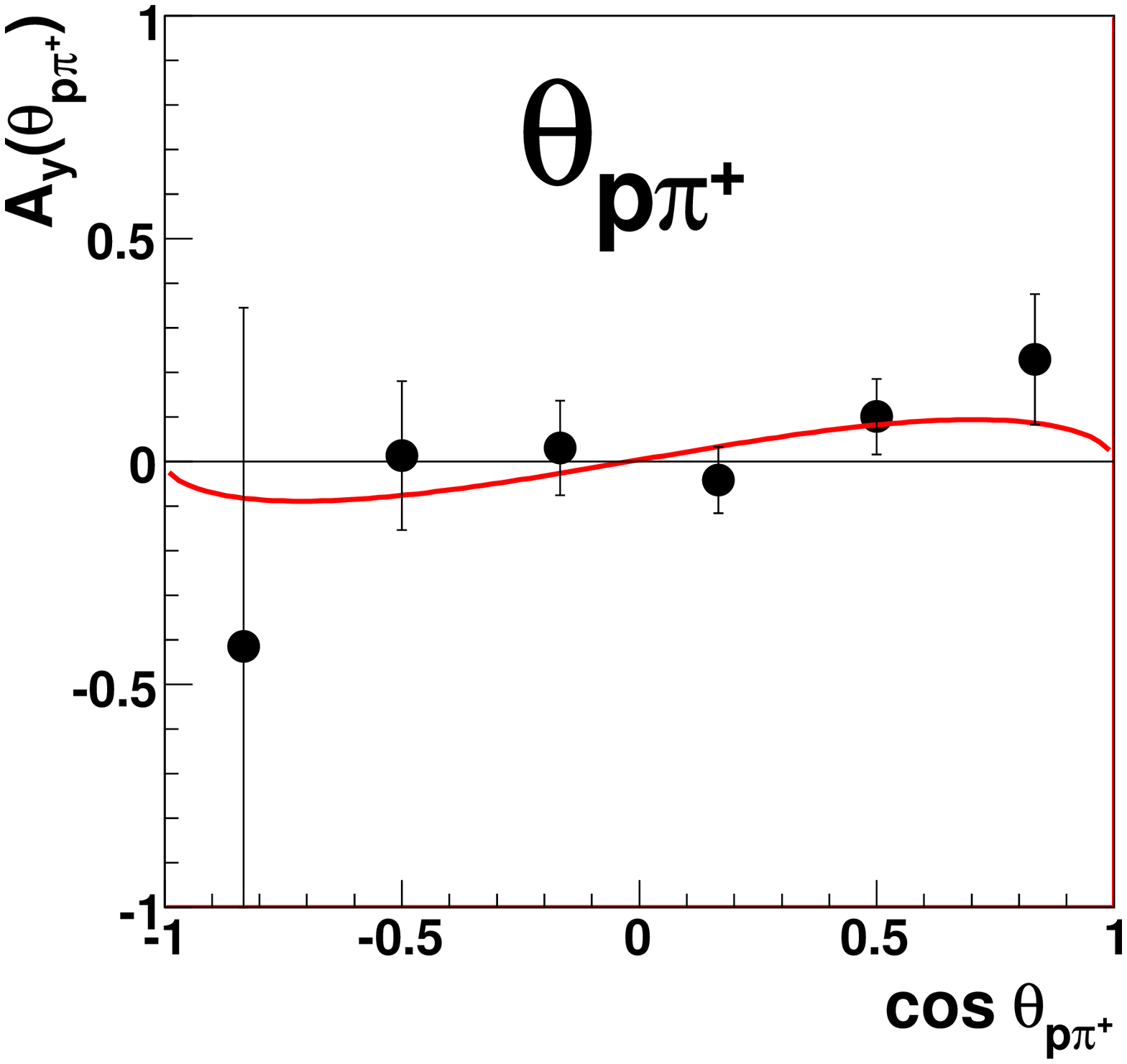}
\includegraphics[width=10pc]{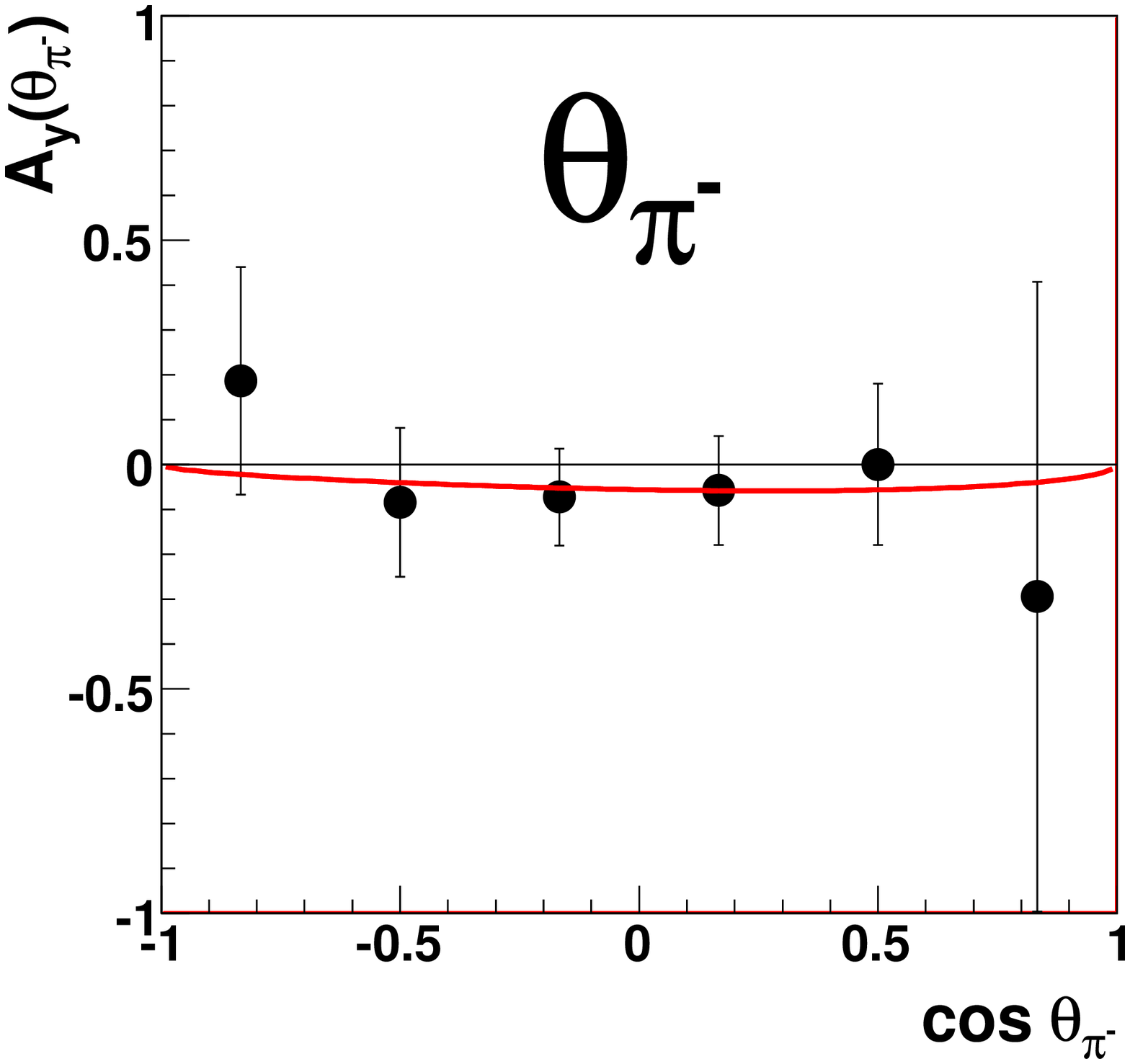}
\includegraphics[width=10pc]{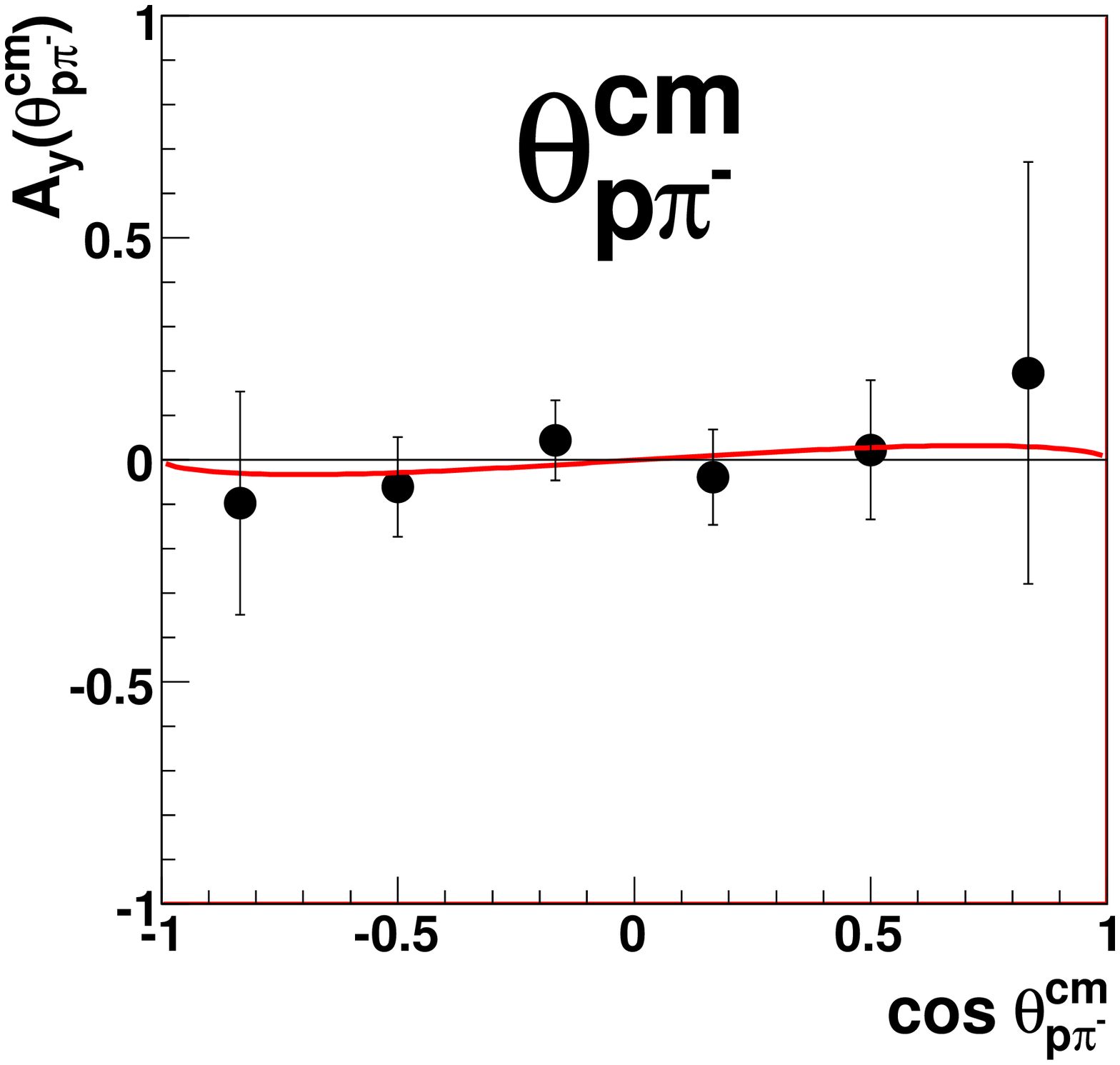}
\end{center}
\caption{Analyzing power distributions for the polar angles
  $\Theta_{\pi^+}^{cm}$, $\Theta_{\pi^-}^{cm}$, $\Theta_{p\pi^+}^{cm}$ and
  $\Theta_{p\pi^-}^{cm}$ at   $T_p$ = 800 MeV, i.e. separated for positively
  and negatively charged pions. The curves show a fit according to
  eq.(7).  }
\end{figure}

\begin{figure}
\begin{center}
\includegraphics[width=10pc]{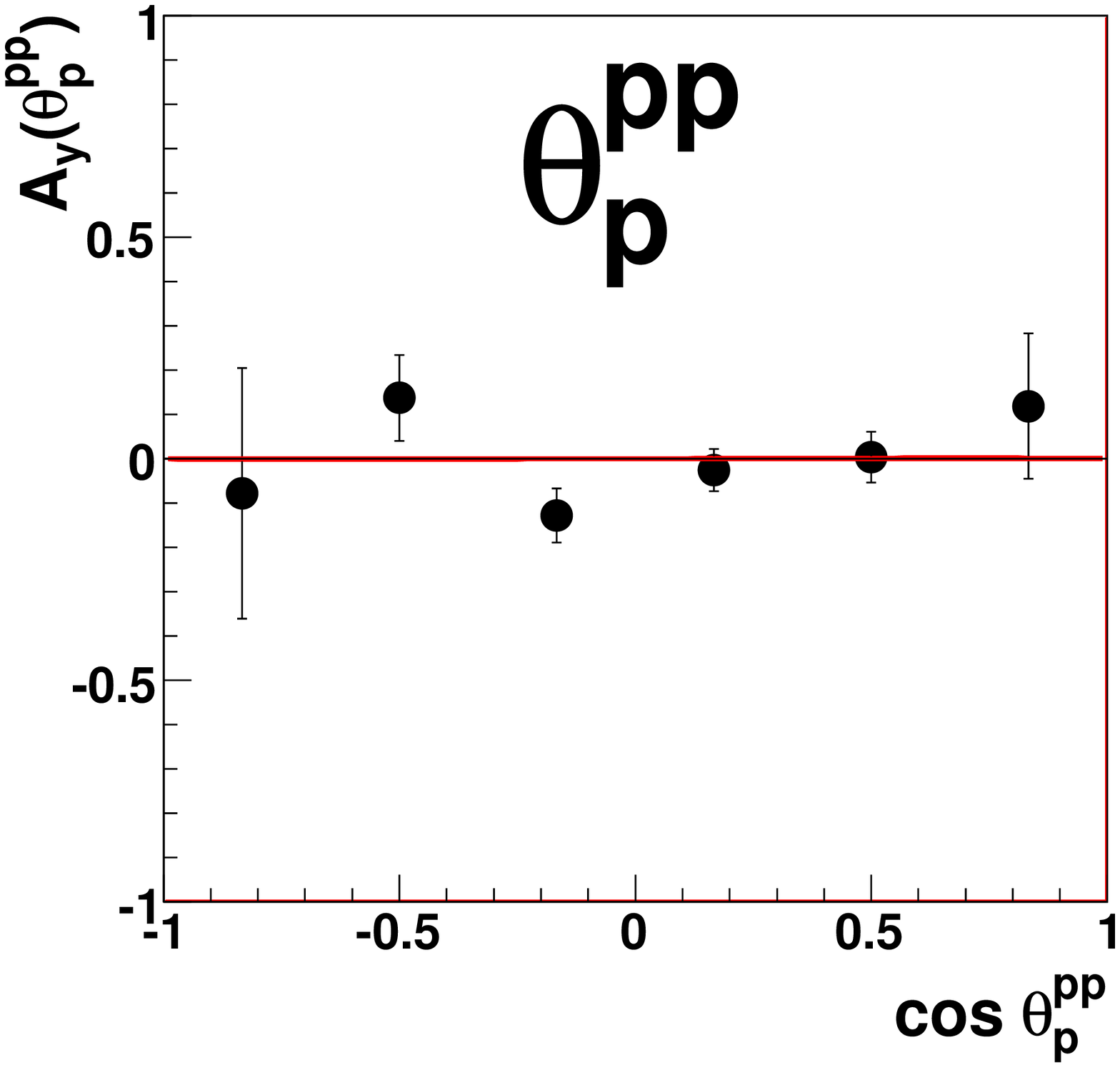}
\includegraphics[width=10pc]{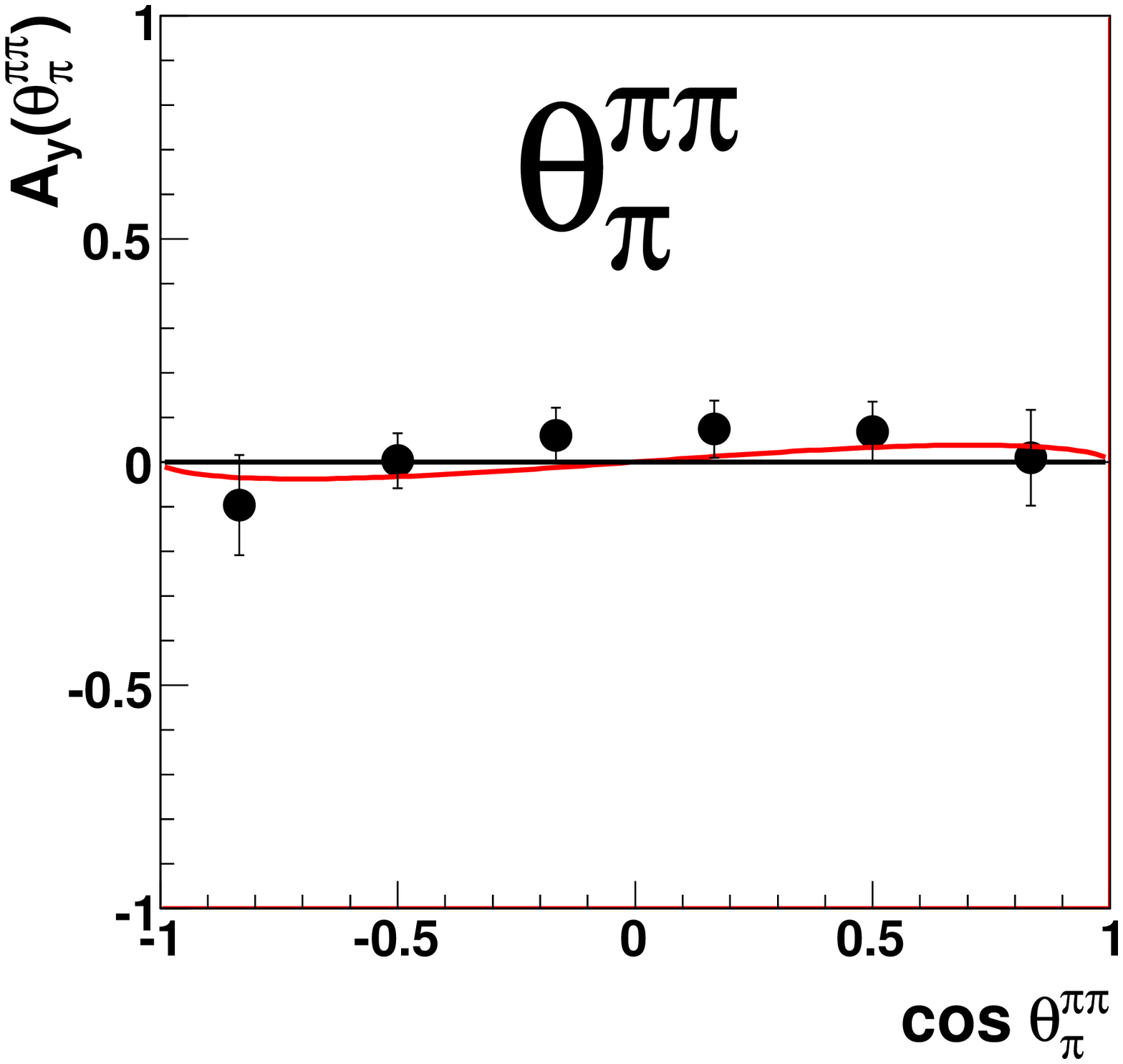}
\end{center}
\caption{Analyzing power distributions for the polar angles $\Theta_{p}^{pp}$,
  $\Theta_{\pi}^{\pi\pi}$ in the $pp$ and $\pi\pi$ subsystem (Jackson frame) at
  $T_p$ = 800 MeV. The curves show a fit according to eqs.(7) and (8),
  respectively.  }
\end{figure}

\section{Discussion of Results}
\label{sec:4}

\subsection{Cross Sections - Unpolarized}
\label{sec:4.1}

The total cross sections obtained in this work are shown in Fig. 9 together
with data from previous work \cite{cverna,coch,brunt,shim,dakhno,gat,WB} and
theoretical calculations of Ref \cite{luis}. At $T_p$ = 750 MeV our value
agrees well with those obtained at CELSIUS \cite{WB,JP}. At $T_p$ = 800 MeV
our value is consistent with the estimate of Ref. \cite{cverna} and follows
also the trend given by the very recent bubble chamber results from Gatchina
\cite{gat}, which correct the old Gatchina results \cite{dakhno} for the total
cross sections.

The theoretical calculations of Ref. \cite{luis} are compatible with the data
from LAMPF \cite{cverna}, CELSIUS \cite{WB,JP}, Gatchina \cite{gat} and
COSY-TOF for $T_p \leq$ 800 MeV only, if set I (see Ref. \cite{luis}) for the
Roper resonance parameters is used - and  if the $pp$ final state interaction 
(FSI) is neglected (lower dashed curve in Fig. 9). Taking into account the FSI
(solid curves)
the theoretical calculations overestimate the cross sections in the
near-to-threshold region by as much as an order of magnitude.


For the discussion in particular of the data taken at $T_p$ = 800 MeV, we
compare our data in Figs. 10 - 13 with phase phase distributions denoted by
the hatched areas in the figures, with theoretical calculations of
Ref. \cite{luis} (Fig. 13 only) as well as with simplified calculations of
Roper excitation and decay according to the ansatz for the Roper amplitude 

\begin{center}
$A \sim 1 + c \bf{k_1} * \bf{k_2} (3D_{\Delta^{++}} +
D_{\Delta^0}),~~~~~~~~~~~~~~~~~~(6)$ 
\end{center}

which multiplies the expressions for $\sigma$ exchange, FSI and $N^*$
propagator (see Refs. \cite{WB,JP}) and where $D_\Delta$ denotes the $\Delta$
propagator. 
The first term in eq. (6) stands for the Roper decay into the $N\sigma$
channel. The second term gives the Roper decay via the $\Delta$
resonance, i.e. $N^* \to \Delta\pi \to N\pi\pi$, where the scalar product of the
pion momenta $\bf{k_1}$ and $\bf{k_2}$ stands for the double p-wave emission of
the pions in this process.  Whereas the first term provides a phase-space like
behavior of the observables, the second term is proportional to the cosine
of the $\pi\pi$ opening angle in the cms and hence provides
both a change in the slope of the  $\delta_{\pi\pi}$ spectrum and a shift
of the $M_{\pi\pi}$ distribution towards higher masses. From this apparent
signature in these observables the parameter c can be fixed easily by 
corresponding data.
 
The coefficient c contains the branching of the decays $N^* \to N
\sigma$ and $N^* \to \Delta \pi$. Adjustment of c to the CELSIUS data
at $T_p$ = 750 and 775 MeV \cite{JP} resulted in a branching ratio of
$R = \Gamma_{N^* \to \Delta \pi \to N\pi\pi} / \Gamma_{N^* \to
  N\sigma} $ between the two-pion decays of the Roper resonance of $R$
= 3(1), if the pole of the Roper resonance is assumed to be at its nominal
value of 1440 MeV - in good agreement with the PDG value of 4(2) \cite{PDG},
but more precise. However, if the pole is taken to be
at 1371 MeV, which meanwhile is considered to be more correct
\cite{PDG,piSAID,BES,morsch,sarant}, then this ratio changes to $R$~=~1.0 (1).
Calculations with this value of c are shown in Figs. 10 - 12
by the solid lines. They give a qualitative description of the main
features in the differential data. At a closer look, however,
they provide a shift in the $M_{\pi\pi}$ spectrum and a slope in the
$\delta_{\pi\pi}$ spectrum, which is somewhat too large. This suggests
that the $N^*\to \Delta\pi$ decay is even weaker than the $N^* \to
N\sigma$ decay - as was also obtained in a very recent analysis of photon-
and pion-induced reactions \cite{sarant}. The result obtained here is also in
agreement with a recent analysis of $pp
\to pp\pi^0\pi^0$ data taken at CELSIUS-WASA \cite{TS}.

Fig. 13 shows the full calculations of Ref.\cite{luis}, which also include
reaction processes other than just the Roper excitation. The dotted line
corresponds to calculations with the same branching for the $N^* \to N\sigma$
and $N^*\to \Delta\pi$ decay processes as used with the simplified ansatz
(6). We see by comparison of Figs. 11 and 12 with Fig. 13 that full and
simplified calculations give nearly the same result, which means that the
terms neglected in the ansatz (6) are of minor importance. In fact, if we
now reduce the  $N^*\to \Delta\pi$ amplitude by a factor of two in the
calcultions of Ref.\cite{luis} -- as done in Ref. \cite{TS} -- then we get
full agreement with the data.

\subsection{Analyzing Powers}
\label{sec:4.2}

The analyzing powers taken at $T_p$ = 750 MeV shown in Fig. 14 are essentially
compatible with zero within their large statistical uncertainties. Because of
that we 
do not discuss them here in more detail and concentrate on the data at $T_p$ =
800 MeV, which have much better statistics. We only note in passing, that the
trend in the analyzing powers at $T_p$ = 750 MeV is just opposite in sign to
that in the data at $T_p$ = 800 MeV. The reason for it is not known to us. 

At $T_p$ = 800 MeV the data displayed in Figs. 15 - 17 exhibit significant
analyzing powers with values up to $A_y$~=~0.3. Whereas the analyzing powers in 
the $pp$ and $\pi\pi$  subsystems are very small and compatible with zero,
they are quite substantial in the overall cms for $\Theta_{\pi}^{cm}$ and
$\Theta_{\pi\pi}^{cm}$. The 
analyzing power of the protons in the overall cms, i.e. $A_y(\Theta_p^{cm})$, is
again small.

The analyzing powers for $\Theta_{\pi}^{cm}$ and  $\Theta_{\pi\pi}^{cm}$ are of
course not decoupled.
 Since for the latter the analyzing power is largest, the
source for polarization producing effects is expected to be connected with this
observable.

From general considerations \cite{hom,han} of meson production in $NN$ collisons
the analyzing powers in the overall cms has the following $\Theta$
dependences (see appendix for more details): 
\begin{center}
$A_y$ = $ a~sin(\Theta)~+~b~sin(2\Theta),~~~~~~~~~~~~~~~~~~~(7)$
\end{center}
whereas in the $pp$ subsystem we have only a $2\Theta$ dependence: 
\begin{center}
$A_y$ = $ b'~sin(2\Theta).~~~~~~~~~~~~~~~~~~~~~~~~~~~~~~~~~~~(8)$
\end{center}
A more detailed description within the partial wave concept is given in the
appendix. 
The solid lines in Figs. 15 - 17 show fits to the data with eqs. (7) and (8),
respectively. As
one can see from Figs. 15 - 17, the
coefficients $a$ and $b$ are only significantly different from zero for 
$\Theta_{p}^{cm}$, $\Theta_{\pi}^{cm}$ and  $\Theta_{\pi\pi}^{cm}$
distributions. In addition the latter ones are essentially described by a
$sin(\Theta)$ dependence with the $sin(2\Theta)$ dependence being of minor
importance or negligible.

Finally we look into possible differences between $\pi^+$ and $\pi^-$ angular
distributions by exploiting the delayed pulse technique to identify the
$\pi^+$ particle in an event. That way we obtain the data distinguished
between $\pi^+$ and $\pi^-$ observables in Fig. 16. Though the statistics
gets worse due to the finite $\pi^+$ identification efficiency, we find the
analyzing powers for the $\pi^+$ observables in $\Theta_{\pi^+}^{cm}$ and 
$\Theta_{p\pi^+}^{cm}$ to be systematically larger than for those for the
corresponding $\pi^-$ observables. Such a difference indicates a significant
contribution of isovector $\pi\pi$ pairs.

From this and the requested sine dependence of the analyzing power we learn
that the driving term for finite analyzing powers must be
connected primarily with the $\pi^+$ momentum and be of the form 
\begin{center}
$\mathbf{\sigma} \cdot (\mathbf{q}~\times~\mathbf{k_{\pi^+}} ),~~~~~~~~~~~~~~~~~~~~~~~~~~~~~~~~~~~~~~~~(9)$
\end{center}
where $\bf{q}$ and $\bf{k_{\pi^+}}$ are momentum transfer between the nucleons
and 
the $\pi^+$ momentum, respectively, and $\sigma$ is the Pauli spin
operator. Since the momentum transfer close to 
threshold is approximately given by the initial momentum in the overall cms,
we arrive with such an operator at the observed sin($\Theta_{\pi^+}^{cm}$)
dependence.

Since to lowest order
spinflip terms like (9) are connected with p-waves, whereas nonspinflip terms
are connected with s-waves, the analyzing power arises to lowest order from an
interference of S- and P-waves in the entrance channel. The dominating reaction
process at $T_p \approx$ 800 MeV is via the Roper excitation and hence fed by
the  $^1S_0$ partial wave in the entrance channel. 
Processes involving the operator of eq. (9) evolve via the $^3P_1$ partial
wave.  
From inspection of the formulas given in Ref.\cite{luis} we see that neither
Roper nor $\Delta\Delta$ excitation and decays provide such an operator. The
only process, which provides such an operator in the theoretical
investigations of Ref. \cite{luis}, is single $\Delta$ excitation and decay in
combination with pion rescattering on the other vertex as given by the graphs
(10) and (11), Fig.1 of Ref. \cite{luis}, or graph (9), which gives a
successive excitation of a $\Delta$ on a single nucleon . However, in these
calculations such graphs give only tiny contributions to the cross section. On
the other hand the analyzing power is an ideal tool to reveal such small
amplitudes by their interference with large amplitudes. Hence it appears very
interesting to check by realistic calculations, whether such graphs could be
the origin of the finite analyzing powers we observe. As pointed out above,
the calculations of Ref. \cite{luis} are done in the plane wave limit, i.e.,
are not valid for 
polarization observables. Hence an improved calculation, which includes
initial state interactions, would be of great help to understand the
polarization data on a quantitative basis. 

Another interesting aspect could be
the consideration of a possible influence of the $\Delta(1600)$
resonance. Though its resonance energy is as high as that of the
$\Delta\Delta$ excitation, its unusually large width of $\approx$ 350 MeV
\cite{PDG} is in favor of taking part in the two-pion production already close
to threshold. Indeed, as has been noted already in Ref. \cite{jan}, the
isotensor two-pion production amplitude is observed to be unusually large. And
the $\Delta(1600)$ resonance is the only resonance candidate to
contribute to the isotensor amplitude near threshold in addition to the
$\Delta\Delta$ process.

The analyzing power arises from the interference of spinflip and
non-spinflip terms in the reaction amplitude, whereas the unpolarized cross
section is just the sum of their moduli squared. If we denote the non-spinflip
term by g and the strength of the spinflip term which multiplies the operator
(9) by h, then we have 
\begin{center}
$\sigma A_y$ = $ -2 Im(g^*h)~k_{\pi^+}
~sin(\Theta_{\pi^+}^{cm})~~~~~~~~~~~~~~~~~~~~~(10)$ 
\end{center}
and

\begin{center}
$\sigma$ = $ |g|^2 + |h|^2 k_{\pi^+}^2  ~sin^2(\Theta_{\pi^+}^{cm})
\leq |g|^2 ( 1 + A_y^2 / 4 ),~~~~~(11)$
\end{center}

i.e., a 30$\%$ effect in the analyzing power produces only a 2$\%$ effect of
non-isotropy in
the differential cross section. Such an effect is not observable in the cross
section with the statistics accumulated in this measurement.

\section{Summary}
\label{sec:5}
The experimental results presented in this work constitute the first exclusive
measurements of $\pi^+\pi^-$ production in $NN$ collisions at $T_p$ = 800 MeV
and provide the first polarization data at all for this production
process. The unpolarized differential cross sections support the findings of
previous analyses that Roper excitation and decay  is the dominant  two-pion
production process. The data favor the direct decay of the Roper resonance
into the $N\sigma$ channel over its decay via  the
$\Delta$ resonance. This in turn supports the breathing
mode nature of the Roper resonance.

The observation  of sizable analyzing powers in combination with the
fact that neither Roper nor $\Delta\Delta$ processes contribute to substantial 
analyzing powers, point to a  small amplitude of different nature, which
interferes with the strong Roper amplitude. Candidates for such a small
amplitude could be $\Delta$ excitation processes, where on the one nucleon
vertex a $\Delta(1232)$ or$\Delta(1600)$ is
excited followed by its transition $\Delta \to \Delta\pi \to N\pi\pi$, which
would be the first time that such a process has been sensed in $NN$
collisions.

\begin{acknowledgement}

This work has been supported by BMBF, DFG (Europ. Gra\-duiertenkolleg
683) and COSY-FFE. We acknowledge valuable discussions with Eulogio Oset,
Murat Kaskulov and Alexander Sibirtsev. We are particularily indebted to Luis
Alvarez-Ruso, who made his computer code available to us. 
\end{acknowledgement}

\section{Appendix}

In this appendix the general structure of the amplitudes is derived.
Since none of the observables shows a stricing energy dependence
and the energy is still not too far above threshold it appears
justified to work with the lowest partial waves only. Hence 
we will only derive the angular structures of differential
cross section and analyzing power up to order $Q^2$, were $Q$ 
is any of the momenta in the final state as defined in Fig. \ref{koorsys}.
\begin{figure}\begin{center}
\psfrag{a}{$\vec k\, '$}
\psfrag{b}{$\vec q\, '$}
\psfrag{c}{$\vec p\, '$}
\psfrag{d}{$\vec p$}
\includegraphics[height=4cm]{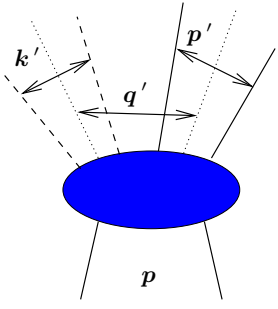} \end{center}
\caption{Definition of the coordinate system. Dashed (solid) lines
denote pions (nucleons).}\label{koorsys}
\end{figure}

Before we can identify the partial waves that are relevant given the
assumption formulated above, we need to briefly comment on the
relevant selection rules. The Pauli principle enforces that a proton
pair of spin 0 (1) has even (odd) parity.  As a consequence, for
unpolarized observables $pp$ $S$--waves and $pp$ $P$--waves never
interfere. Since in the experiment presented the polarization of the
outgoing protons is not measured, no interferences of $pp$ $P$-- and
$S$--waves in the final state are possible. We therefore only need to
consider $pp$ $P$--waves in the final state in combination with
$s$--waves for the other two subsystems ($\pi\pi$ and $\pi\pi$
relative to $pp$).

  There are 4 partial waves possible with an NN p-wave in the final state ($^3P_j\to
{}^3P_j s[s]$ with $j=0,1,2$ and $^3F_2\to {}^3P_2 s[s]$, where we
used the notation $^{2S+1}L_j\to {}^{2S'+1}L'_{j'} \bar l[l]$, with $S$,
$L$, $j$ ($S'$, $L'$, $j$) for spin, angular momentum, and total
angular momentum of the $pp$ pair in the initial (final) state, 
$l$ for the angular momentum of the pion pair, and $\bar l$ for
the angular momentum of the pion pair relative to the $pp$ pair.
It is straight forward to see that only 3 independent operator structures
are possible
and we define
$$
{\cal M}^P=d_1(\vec S\cdot \hat p)(\vec S\, '\cdot \vec p\, ')+
d_2(\vec S\cdot \hat p\, ')(\vec S\, '\cdot \vec p)
+d_3(\vec S\cdot \vec S\, ')(\hat p \cdot \vec p\, ') \ .
$$
The partial wave amplitudes are linear combinations of the
$d_i$.

 Terms of at most quadratic dependence on the final momenta
in combination with $pp$ $S$--waves in the final state can
emerge from overall $S$--waves interfering either with
themselves, $p$--waves in the subsystems or $d$--waves in the 
subsystems. We therefore need to consider in total 6 partial
waves for this part that we parametrize as
\begin{eqnarray*}
{\cal M}^S=a&+&b_1(\vec S\times \hat p)\cdot \vec q\, '
+b_2\left((\vec q\, '\cdot \hat p)^2-\frac13\vec q\, '{}^2\right)+ \\
&+&c_1(\vec S\times \hat p)\cdot \vec k\, '
+c_2(\vec q\, '\cdot \vec k\, ')+ \\
&+&c_3\left((\hat p\cdot \vec q\, ')(\hat p \cdot \vec k\, ')-\frac13(\vec q\, '\cdot \vec k\, ')\right) \ ,
\end{eqnarray*}
where the $a$ term corresponds to $^1S_0\to {}^1S_0s[s]$, 
the $b_1$ term to $^3P_1\to {}^1S_0p[s]$, 
the $b_2$ term to $^1D_2\to {}^1S_0d[s]$, 
the $c_1$ term to $^3P_1\to {}^1S_0s[p]$, 
the $c_2$ term to $^1S_0\to {}^1S_0p[p]$, and 
the $c_3$ term to $^1D_2\to {}^1S_0p[p]$. 
The full matrix element is given by
$$
{\cal M}={\cal M}^S+{\cal M}^P \ .
$$ 

Using the formulas of chapter 4.3 of Ref.~\cite{han} it is straightforward
to calculate both unpolarized cross sections as well as analyzing
powers from the expressions given above. We find (dropping terms more
than quadratic with respect to the final momenta)
\begin{eqnarray*}
4\sigma_0 = |a|^2 &+& |b_1|^2q'\,^2\sin(\theta_{q'})^2
+|c_1|^2k'\,^2\sin(\theta_{k'})^2+ \\
&+& 2\mathrm{Re}\{ a^* b_2q'\,^2\left(\cos (\theta_{q'})^2-\frac{1}{3}\right)+\\
&+& \frac{2}{3} a^* c_3q'k'\cos (\theta_{q'})\cos(\theta_{k'})+\\ 
&+&\left(a^*c_2-\frac{1}{3} a^*c_3+b_1^*c_1\right)k'q' \\
&&\sin (\theta_{k'})\sin (\theta_{q'})
\cos(\phi_{q'}-\phi_{k'})\} \\
&+&p'\,^2\left(|d_1|^2+|d_2|^2\right)+ \\
&+&p'\,^2\cos(\theta_{p'})^2 \left (3|d_3|^2 +
2\mathrm{Re}\left\{d_1^*d_2+d_1^*d_3+d_2^*d_3\right\}\right) \\
4\sigma_0A_y= &-&\mbox{Im}\left\{\left(d_1^*d_2+d_1^*d_3-d_2^*d_3\right)p'\,^2
\sin (2\theta_{p'})\cos(\phi_{p'})\right.
\\
&&\left.+2b_1^*aq'\sin(\theta_{q'})\cos(\phi_{q'})
+2c_1^*ak'\sin(\theta_{k'})\cos(\phi_{k'})\right\} \ .
\end{eqnarray*}
Note that, as a consequence of the fact that two amplitudes with a spin singlet
entrance channel can not interfere in the analyzing power, the expression
for the polarization observable has a much simpler structure.

}

\end{document}